\documentclass{aastex62}
\usepackage{amsmath}
\usepackage{amssymb}
\usepackage{graphicx}
\graphicspath{{./figures/}}
\usepackage{ dsfont }
\usepackage{enumitem}
\usepackage{aas_macros}

\usepackage{color}

\renewcommand{\abstractname}{}
\newcommand{\eq}[1]{\begin{equation}\begin{split}#1\end{split}\end{equation}}

\newcommand{\eal}[1]{\begin{align}#1\end{align}}
\newcommand{\hb}[1]{\hat{\boldsymbol{#1}}}

\newcommand{\bs}[1]{\boldsymbol{#1}}

\newcommand{\red}[1]{{#1}}

\shorttitle{Exciting mutual inclination with a stellar companion}
\shortauthors{Xu and Fabrycky}

\begin{document}
\title{Exciting mutual inclination in planetary systems with a distant stellar companion: the case of Kepler-108}

\correspondingauthor{Wenrui Xu}
\email{wenruix@princeton.edu}

\author[0000-0002-9408-2857]{Wenrui Xu}
\affiliation{Department of Astrophysical Sciences, Princeton University, 4 Ivy Lane, Princeton, NJ 08544, USA}
\author[0000-0003-3750-0183]{Daniel Fabrycky}
\affiliation{Department of Astronomy and Astrophysics, University of Chicago, 5640 S. Ellis Ave., Chicago, IL 60637, USA}

\begin{abstract}
We study the excitation of mutual inclination between planetary orbits by a novel secular-orbital resonance in multiplanet systems perturbed by binary companions which we call ``ivection". The ivection resonance happens when the nodal precession rate of the planet matches a multiple of the orbital frequency of the binary, and its physical nature is similar to the previously-studied evection resonance. Capture into an ivection resonance requires encountering the resonance with slowly increasing nodal precession rate, and it can excite the mutual inclination of the planets without affecting their eccentricities.
{We discuss the possible outcomes of ivection resonance capture, and we use simulations to illustrate that it is a promising mechanism for producing the mutual inclination in systems where planets have significant mutual inclination but modest eccentricity, such as Kepler-108. We also find an apparent deficit of multiplanet systems which would have nodal precession period comparable to binary orbital period, suggesting that ivection resonance may inhibit the formation or destablize multiplanet systems with external binary companion.}
\end{abstract}

\section{Introduction}
Thousands of exoplanets have been discovered so far, and nearly half of exoplanet systems host multiple observed planets \citep{Burke14}. 
However, among all these systems, only three are observed to have significant (above a few degrees), measured mutual inclinations to date.\footnote{There are also a few circumbinary planets that are slightly inclined with respect to the binary; here we only consider planets around a single star.} Kepler-419 b and c have a marginally detected mutual inclination of ${9^\circ}^{+8}_{-6}$, which is modest given the high eccentricities of the planets \citep{Dawson12}. The other two systems, Kepler-108 and Upsilon Andromeda, both have significant mutual inclination and modest eccentricity. Kepler-108 b and c have a mutual inclination of ${24^\circ}_{-8}^{+11}$, with eccentricities $0.135^{+0.11}_{-0.06}$ and $0.13\pm 0.02$ respectively \citep{MillsFabrycky17}. Upsilon Andromeda c and d have mutual inclination of $30^\circ\pm 1^\circ$ and eccentricities $0.245\pm 0.006$ and $0.316\pm0.006$ respectively \citep{McArthur10}.\footnote{There is one other confirmed planet (Upsilon Andromeda b) and one unconfirmed planet (Upsilon Andromeda e) in the system, but their masses are small compared to Upsilon Andromeda c and d and should not affect the evolution of Upsilon Andromeda c and d.}

The large mutual inclination and small eccentricities of Kepler-108 are difficult to explain, if the planets were formed in a coplanar configuration.
Exciting the mutual inclination via scattering with another planet is possible,
but producing such low eccentricity requires some fine-tuning since planet-planet scattering tends to produce eccentricities that are comparable to or larger than the mutual inclination (in radians) \citep{Chatterjee08}. 
The origin of the mutual inclination may also be due to a binary companion:
Kepler-108 has a binary companion with sky-projected separation of 327 AU. (The eccentricity and semi-major axis of the binary remain unknown.)
{However, given the large separation, the gravitational perturbation of the binary companion would be too weak to affect the evolution of the planets on a dynamical timescale. In addition, since the system hosts two relatively massive planets, the precession of the planets due to perturbation from each other completely suppresses secular inclination excitation via Lidov-Kozai oscillation \citep{MillsFabrycky17}.}

Although known mechanisms are having difficulty producing the mutual inclination of Kepler-108, the similarity of the planets' nodal precession rate and the binary's orbital frequency suggests that the inclination may be related to a secular-orbital resonance between the planets and the binary \citep{MillsFabrycky17}.
In this paper, we aim to provide a plausible explanation for the mutual inclination of Kepler-108 using a novel resonance between the nodal precession of the planets and the orbital motion of the binary. This new resonance is similar to evection resonance, a resonance between the apsidal precession of the planet and the orbital motion of the binary that can excite the eccentricities of planets in a multiplanet system \citep{ToumaSridhar15}.\footnote{Evection resonance is originally studied in the context of lunar evolution perturbed by the sun \citep{ToumaWisdom98}, and can also be applied to exomoons \citep{Spalding16} and circumbinary planets \citep{XuLai16}.} This new resonance we identify is named ``ivection" resonance, to signify that it is highly similar to evection resonance but excites the inclination instead of eccentricity of the planet.\footnote{{\citet{ToumaWisdom98} discussed a similarly named ``eviction" resonance, which is an inclination-eccentricity resonance with resonant term $\propto e^2I$. This resonance can also excite inclination, but only when the perturbed body has finite initial eccentricity. It is different from the ivection resonance we discuss here.}}

Our discussion is organized as follows: In Section \ref{sec:ivection}, we introduce the mechanism of ivection resonance and derive the Hamiltonians of different types of ivection resonance, namely first and second-order ivection resonance (for a near-circular binary) and eccentric ivection resonance (for a very eccentric binary). Then, in Section \ref{sec:outcome}, we study ivection resonance capture during planet migration, discussing the requirements (especially on the rate of migration) and possible outcomes of resonance capture. Section \ref{sec:K108} applies results from the previous sections to study the formation of Kepler-108, and we reproduce the observed eccentricities and mutual inclination with numerical simulation. In Section \ref{sec:application} we discuss the importance of ivection resonance in other exoplanet systems.
We conclude with disussions of a few topics related to this study in Section \ref{sec:discussion} and a summary of the main results in Section \ref{sec:summary}.

\section{Ivection resonances}\label{sec:ivection}

Consider two planets with mass $m_1,m_2$ (subscript 1 denotes the inner planet) with initially coplanar orbit around their host (with mass $M_\star$).
The planets are perturbed by a distant binary companion with mass $M_B$ and constant eccentricity $e_B$.
{The Hamiltonian governing the secular evolution of the planets is given by}
\eq{
H_{\rm sec} = -\left\langle\frac{Gm_1m_2}{|\bs r_1 - \bs r_2|}\right\rangle - GM_Bm_1\left\langle\frac{1}{|\bs r_B-\bs r_1|}-\frac{\bs r_B\cdot \bs r_1}{r_B^3}\right\rangle - GM_Bm_2\left\langle\frac{1}{|\bs r_B-\bs r_2|}-\frac{\bs r_B\cdot \bs r_2}{r_B^3}\right\rangle.\label{eq:Ham_full}
}
{Here $\bs r_1,\bs r_2$ and $\bs r_B$ are the position of the planets and the binary companion with respect to the central star, and $\langle\rangle$ denotes averaging over the orbits of the two planets.
In what follows, we assume that the two planets are not close to any mean motion resonance, and that the planet orbits remain circular.
The latter is a good approximation if the initial eccentricity is small, since the secular coupling between eccentricity and inclination is at least fourth order \citep{MurrayDermott99}.}

{The first term of \eqref{eq:Ham_full} describes the secular coupling between the two planets, which produces a nodal precession but does not directly excite mutual inclination.
Let $I$ be the mutual inclination of the two planets and $\Omega$ be the longitude of ascending node of the outer planet, defined with respect to the invariable plane of the two planets (i.e. the plane aligned with their total angular momentum).
Note that on long timescales the invariable plane is not fixed, but precesses around the binary angular momentum.}
The nodal precession rate $d\Omega/dt$ due to the planets perturbing each other is given by (see derivation in Appendix \ref{a:Ham})
\eq{
\frac{d\Omega}{dt} = \dot\Omega_0 \left[1 + \frac 12 \left(\frac{f_8}{f_3}-\frac{\beta}{(1+\beta)^2}\right)I^2\right].\label{eq:dOmega_sec}
}
with
\eq{
\dot\Omega_0\equiv \frac 12 n_2\mu f_3,~~~
\mu\equiv \frac{m_1}{M_\star}(1+\beta),~~~\beta\equiv\frac{m_2}{m_1}\left(\frac{a_2}{a_1}\right)^{1/2}.
}
$f_3,f_8$ are $\mathcal O(1)$ functions of $\alpha = a_1/a_2$ given in Appendix B of \citet{MurrayDermott99} evaluated at $j=0$. Note that $f_3<0, f_8>0$; therefore nodal precession is retrograde ($\dot\Omega<0$) and the precession rate $|\dot\Omega|$ decreases as $I$ increases.
In the limit of small $\alpha$, $-f_3\approx f_8\approx 1.5\alpha^2$ and the period of precession is approximately given by
\eq{
\frac{2\pi}{|\dot\Omega_0|} \approx 1300\rm{~yr}\times\left(\frac{P_2}{1\rm{~yr}}\right)\left(\frac{P_1}{P_2}\right)^{-4/3}\left(\frac{\mu}{10^{-3}}\right)^{-1}.
}
Here $P_1,P_2$ are the orbital period of the planets. This expression can also be used as a crude estimate for $\dot\Omega_0$ when $a_1$ and $a_2$ are comparable (for more accuracy, see \citealt{BaileyFabrycky2020}).

{The other two terms in \eqref{eq:Ham_full} describe the perturbation from the binary. Such perturbation is typically small, but it is nontrivial when the precession rate of the planets becomes commensurate with some integer multiple of the orbital frequency of the binary.
We call this secular-orbital resonance the ivection resonance.}
The name ``ivection" is derived from from evection resonance, which is the resonance between the apsidal precession and the orbital frequency of some distant perturber \citep{ToumaWisdom98,ToumaSridhar15}. The replacement of e by i signifies that ivection resonance affects the inclination (instead of eccentricity) of the system.
Ivection resonances happen when (assuming $a_1/a_2$ is of order unity, so that $|f_3|$ is also of order unity)
\eq{
n_B \sim |\dot\Omega| \sim n_2\mu,~~~ a_B\sim a_2\mu^{-2/3}.
}
For planets with smaller mass or longer period, the binary has to be further away for ivection resonance to happen.

{In the remaining of this section, we analyze the perturbation from the binary to define different types of ivection resonances and their corresponding resonant angles in \S\ref{sec:Phi_res}, and construct an effective Hamiltonian for each type of ivection resonance in \S\ref{sec:Ham}.}

\subsection{Resonant perturbation from binary}\label{sec:Phi_res}
In this subsection we study the secular perturbation from the binary.
For simplicity, we only consider two limiting cases: when the binary has zero eccentricity ($e_B=0$), and when the binary has very large eccentricity [$(1-e_B)\ll 1$].
{The results for the limiting cases can be interpreted as the lowest-order term of the resonant perturbation potential expanded in powers of $e_B$ and $(1-e_B)$ respectively.
For intermediate $e_B$ (where expanding in $e_B$ or $(1-e_B)$ is a bad approximation), it is likely that the resonances found for the limiting cases remain qualitatively the same, except that their strengths (amplitudes of the resonant potentials) will be affected by $e_B$.}

\subsubsection{Circular binary}
First consider a circular binary. To quadrupole order in $a_i/a_B$ and lowest order in $m_i/M_\star$, the coupling between a planet and the binary is given by
\eq{
\langle\Phi_{iB}\rangle \equiv -GM_Bm_i\left\langle\frac{1}{|\bs r_B-\bs r_i|} - \frac{\bs r_B\cdot \bs r_i}{r_B^3}\right\rangle = \frac 34 \frac{GM_Bm_ia_i^2}{a_B^3}(\hb{r}_B\cdot \hb{n}_i)^2 + \rm{constant}.
}
Here $\hb{r}_B$ is the direction of the location of the binary, and $\hb{n}_i$ is the direction of angular momentum of planet $i$.
Let $I_i$ and $\Omega_i$ be the inclination of longitude of ascending node of each planet.
Expanding $\langle\Phi_{iB}\rangle$ up to second order in $\sin I_i$ and removing non-resonant terms gives
\eq{
\langle\Phi_{iB}\rangle =& \frac 34 \frac{GM_Bm_ia_i^2}{a_B^3}\left[{-}\sin I_B\frac{1-\cos I_B}{2}\sin I_i\cos(-2\lambda_B+3\Omega_B-\Omega_i)\right.\\
&\left.+\left(\frac{1-\cos I_B}{2}\right)^2\sin^2I_i\sin^2(-\lambda_B+2\Omega_B-\Omega_i)\right].\label{eq:Phi_iB}
}
Here $I_B,\Omega_B$ and $\lambda_B$ are the inclination, longitude of ascending node and mean longitude of the binary, respectively. $I_B$ and $\Omega_B$ are approximately constant.
We ignore the non-resonant terms on the grounds that such terms only change $\dot\Omega_i$ by $\sim n_Ba_i^2/a_B^2$, which is negligible when the system is not too far from ivection resonance. If one wishes to be more exact, the leading order secular planet-binary interaction can be absorbed into terms with the same $I_i$ dependence in the secular planet-planet interaction potential. The form of the resulting Hamiltonian should remain the same.

{The first term in $\langle\Phi_{iB}\rangle$ gives rise to a resonance with resonant angle $(-2\lambda_B+3\Omega_B-\Omega_i)$, which we call a first-order ivection resonance (because the resonant term is $\propto \sin I_i$).
The second term gives rise to another resonance with resonant angle $(-\lambda_B+2\Omega_B-\Omega_i)$, which we call a second-order ivection resonance (because the resonant term is $\propto \sin ^2 I_i$).
Note that $\dot\Omega_i<0,~\dot\lambda_B=n_B>0$ and $\dot\Omega_B\approx 0$.}

Equation \eqref{eq:Phi_iB} shows that the resonant perturbation for the first-order ivection resonance vanishes when the binary is aligned or anti-aligned with the planets ($I_B=0$ or $\pi$), and the resonant perturbation for second-order ivection resonance vanishes when the binary is aligned with the planets ($I_B=0$).
{In general, the resonant perturbation remains qualitatively the same for a prograde and retrograde binary; $I_B$ only affects the strength of the resonant perturbation. We therefore do not discuss prograde and retrograde binaries separately.}

\subsubsection{Very eccentric binary}
The other limit is when the binary is very eccentric, with $(1-e_B)\ll 1$. In this case, the perturbation from the binary produces discrete kicks in planet eccentricity and inclination when the binary passes periastron. Each periastron passage changes inclination by \citep{KobayashiIda01}\footnote{
\citet{KobayashiIda01} derived this relation for planetesimals perturbed by the single passage of a binary with parabolic or hyperbolic orbit. 
The result should also be applicable to elliptic orbits when $e_B\to 1$. We also assume that the change in the vector $(I_i\sin(\Omega-\Omega_B),I_i\sin(\Omega-\Omega_B))$ due to one periastron passage is independent of the initial inclination, which is a relatively good approximation when the initial inclination is small.}
\eq{
&\Delta [I_i\cos(\Omega_i-\Omega_B)] \approx 0,\\
&\Delta [I_i\sin(\Omega_i-\Omega_B)] \approx \frac{3\pi}{8\sqrt{2}}\frac{q}{\sqrt{1+q}}\left(\frac{a_i}{D}\right)^{3/2}\sin 2I_B.\label{eq:kick}
}
Here $q\equiv M_B/M_\star$ is the binary mass ratio, and $D=a_B(1-e_B)$ is the periastron distance.
{The eccentricity of the planet is also perturbed during periastron passage, but the change in eccentricity is smaller than the change in inclination by a factor of $a_i/D$ \citep{KobayashiIda01}. One can safely ignore the eccentricity kicks as long as the apsidal precession frequency is not commensurate with the binary orbital frequency (i.e., the system is far from an ``eccentric evection resonance").}

The inclination kicks are resonant (i.e. the kicks are in the same direction in the frame co-precessing with the planet) {\it iff} the orbital period of the binary is close to an integer multiple of the period of nodal precession.
We call this resonance eccentric ivection resonance (or eccentric first-order ivection resonance), to distinguish it from the ivection resonances for circular binary that we discussed previously.

{To derive the resonant term in the secular binary perturbation potential, we Fourier expand $\langle \Phi_{iB}\rangle$ with respect to the binary mean anomaly and compute the Fourier amplitudes in the limit of small $(1-e_B)$.
When the binary orbital period is $\approx j$ times the planet precession period (with $j$ being a positive integer), 
the resonant term $\langle \Phi_{iB}\rangle$ is given by}
\eq{
\langle \Phi_{iB}\rangle_{\rm res} \approx -\frac{3}{16\sqrt{2}}\frac{GM_Bm_ia_i^2}{a_B^3(1-e_B)^{3/2}}\sin(2I_B)\sin I_i \cos[-j(\lambda_B-\varpi_B)-(\Omega_i-\Omega_B)].
}
{The above expression is a good approximation when $j\ll (1-e_B)^{-3/2}$. The full derivation of this resonant potential is given in Appendix \ref{a:eccentric}.
The same result can also be obtained intuitively by approximating the the discrete kicks in inclination with a continuous forcing corresponding to a potential $\propto \cos[-j(\lambda_B-\varpi_B)-(\Omega_i-\Omega_B)]$.}
The resonant potential of an eccentric ivection resonance is similar to that of a (circular) first-order ivection resonance.

\subsection{Hamiltonian of the system}\label{sec:Ham}

\begin{figure}
\centering
\includegraphics[width=0.8\textwidth]{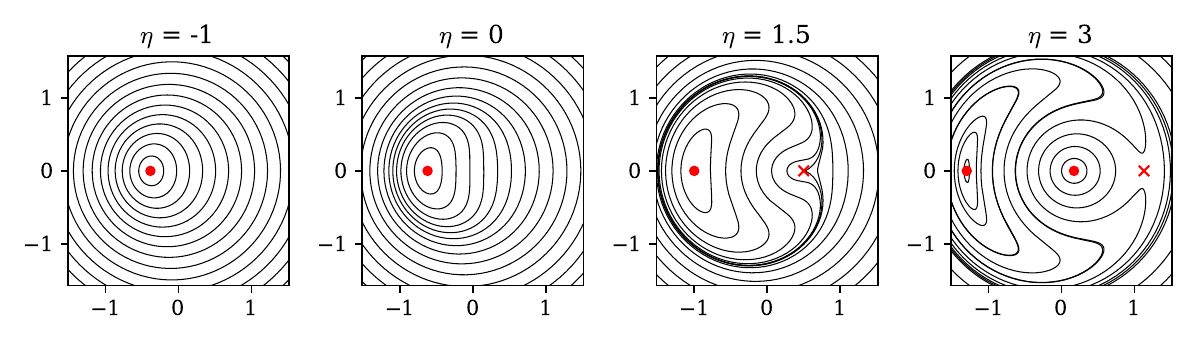}
\includegraphics[width=0.8\textwidth]{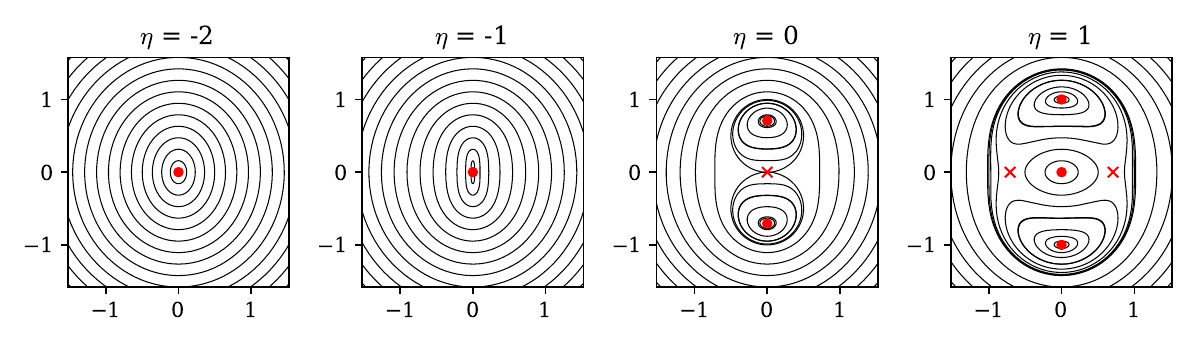}
\caption{Contour of the Hamiltonians \eqref{eq:Ham_1st} and \eqref{eq:Ham_2nd} in $(x,y)$ phase space [see definition at \eqref{eq:xy_def}], at different $\eta$ values.
\textbf{Top panel}: Hamiltonian for first order ivection resonance and eccentric ivection resonance given in \eqref{eq:Ham_1st}. A bifurcation occurs at $\eta=1.5$. The Hamiltonian is similar to that of a first-order MMR.
\textbf{Bottom panel}: Hamiltonian for second order ivection resonance given in \eqref{eq:Ham_2nd}. Two bifurcations occur at $\eta=0$ and 1. The Hamiltonian is similar to that of a second-order MMR.
Stable and unstable fixed points are marked by red dots and crosses respectively. {When the system is captured into resonance, it should librate around the leftmost fixed point for the Hamiltonian in the top panels, and the top or bottom fixed point (with equal probability) for the Hamiltonian in the bottom panels.}}
\label{Ham_contour}
\end{figure}

{For each type of ivection resonance discussed above, we can construct an effective Hamiltonian to describe the evolution of mutual inclination $I$.
In order to do this, we first find the equations of motion for each planet's inclination $I_i$ and resonant angle $\theta_i$ using the secular Hamiltonian $H_{\rm sec}$.
Then we relate $I_i$ and $\theta_i$ to the mutual inclination $I$ and resonant angle $\theta$ (which is $\theta_i$ with $\Omega_i$ replaced by $\Omega$), and find the equations of motion for a pair of variables $(X,Y) \approx (I\cos\theta,I\sin\theta)$ (see their exact definition in Appendix \ref{a:Ham}).
Given $d(X,Y)/dt$, we can build an effective Hamiltonian $H_{\rm eff}(X,Y)$ that gives the correct equations of motion (i.e. $dX/dt = -\partial H_{\rm eff}/\partial Y,~~dY/dt = \partial H_{\rm eff}/\partial X$).
Finally, we scale the Hamiltonian $H_{\rm eff}$ to a dimensionless form $\mathcal H$.}

{For brevity, the full derivation of the effective Hamiltonian is delegated to Appendix \ref{a:Ham}, and below we only summarize the main results.}

\subsubsection{First-order ivection resonance}\label{subsec:1st}
For a first-order ivection resonance, the Hamiltonian of the system can be written in the following dimensionless form:
\eq{
\mathcal H = \eta (x^2+y^2) - (x^2+y^2)^2 - x.\label{eq:Ham_1st}
}
$(x,y)$ are a pair of conjugate variables for $\mathcal H$ given by (for small mutual inclination)
\eq{
&x\approx ({I}/{I_0})\cos\theta, \\
&y\approx ({I}/{I_0})\sin\theta.\label{eq:xy_def}
}
The resonant angle $\theta$ and the characteristic inclination $I_0$ are
\eal{
&\theta = -2\lambda_B+3\Omega_B-\Omega,\\
&I_0= \left[\frac{1}{4} \left(\frac{f_8}{-f_3}+\frac{\beta}{(1+\beta)^2}\right)\right]^{-1/3}\left[\frac 34 \frac{q}{1+q}\frac{1-\cos I_B}{2}\sin I_B \frac{(n_1-n_2)n_B}{n_1n_2}\right]^{1/3}.
}
The time unit for this dimensionless Hamiltonian $T_0$ is given by
\eq{
T_0 = n_B^{-1}\left[\frac{1}{4} \left(\frac{f_8}{-f_3}+\frac{\beta}{(1+\beta)^2}\right)\right]^{-1/3}\left[\frac 34 \frac{q}{1+q}\frac{1-\cos I_B}{2}\sin I_B \frac{(n_1-n_2)n_B}{n_1n_2}\right]^{-2/3}.
}
$\eta$ is a constant characterizing how far the system is from the resonance, and is defined as
\eq{
\eta=\frac 12 \left(\frac{\partial\theta}{\partial\lambda_B}n_B-\dot\Omega_0\right)T_0.
}
For this resonance, $\partial\theta/\partial\lambda_B=-2$.

Physically, the first two terms of $\mathcal H$ represent the secular coupling between the planets, and the last term the resonant perturbation of the binary. For binary mass ratio $q\sim 1$ and planets with similar orbital period (at some timescale $P$), $I_0$ and $T_0$ scale as
\eq{
I_0\sim \mu^{1/3}, ~~T_0\sim \mu^{-5/3}P\sim \mu^{-2/3}P_B.\label{eq:I0_1st}
}
Here $P_B$ is the binary orbital period.

\subsubsection{Second-order ivection resonance}\label{subsec:2nd}
For a second-order ivection resonance, the effective Hamiltonian of the system can be written as
\eq{
\mathcal H = \eta(x^2+y^2)-(x^2+y^2)^2 +y^2,\label{eq:Ham_2nd}
}
with all definitions identical to the previous case except that
\eal{
&\theta = -\lambda_B+2\Omega_B-\Omega,\\
&I_0 = \left[\frac{1}{6}\frac{1+q}{q} \left(\frac{f_8}{-f_3}+\frac{\beta}{(1+\beta)^2}\right)\left(\frac{1-\cos I_B}{2}\right)^{-2}\frac{1+\beta}{1+\beta n_2/n_1}\frac{n_2}{n_B}\right]^{-1/2}\sim \mu^{1/2},\label{eq:I0_2nd}\\
&T_0 = \frac 43 \frac{1+q}{q}\left(\frac{1-\cos I_B}{2}\right)^{-2}\frac{1+\beta}{1+\beta n_2/n_1}\frac{n_2}{n_B^2} \sim \mu^{-2} P.
}
Compared to first-order ivection resonance, the characteristic inclination $I_0$ (which is comparable to the maximum inclination excitation of a non-dissipative system with zero initial inclination) is smaller and the unit time $T_0$ (which characterizes the timescale of libration) is longer, suggesting that second-order ivection resonance is weaker than first-order ivection resonance.

\subsubsection{Eccentric ivection resonance}\label{subsec:ecc}
For an eccentric ivection resonance, the form of the effective Hamiltonian is identical to that of a first-order ivection resonance for circular binary \eqref{eq:Ham_1st} {except that the last term is $\mp x$ for prograde ($I_B<\pi/2$) and retrograde ($I_B>\pi/2$) binary respectively}. The resonant angle $\theta$ and normalizations $I_0$ and $T_0$ are now
\eal{
&\theta = -j(\lambda_B-\varpi_B)-(\Omega-\Omega_B)\\
&I_0 = \left[\frac{j}{8}\left(\frac{f_8}{-f_3}+\frac{\beta}{(1+\beta)^2}\right)\right]^{-1/3}\left[\frac{3}{16\sqrt{2}}\frac{q}{1+q}\frac{n_B(n_1-n_2)}{n_1n_2}(1-e_B)^{-3/2}|\sin 2I_B| \right]^{1/3}\sim \mu^{1/3}(1-e_B)^{-1/2},\label{eq:I0_ecc}\\
&T_0 = n_B^{-1}\left[\frac{j}{8}\left(\frac{f_8}{-f_3}+\frac{\beta}{(1+\beta)^2}\right)\right]^{-1/3}\left[\frac{3}{16\sqrt{2}}\frac{q}{1+q}\frac{n_B(n_1-n_2)}{n_1n_2}(1-e_B)^{-3/2}|\sin 2I_B| \right]^{-2/3}\sim \mu^{-5/3}(1-e_B)P.
}
The large binary eccentricity increases $I_0$ and decreases $T_0$. Note that since $n_B\sim n_2\mu$, $I_0$ is also approximately given by $I_0\sim (a_2/D)^{1/2}$ where $D=a_B(1-e_B)$ is the periastron distance. The stability of the system requires relatively large $D/a_2$, so $I_0$ will never be too large.

\section{Conditions and outcomes of ivection resonance capture}\label{sec:outcome}

\begin{figure}
\centering
\includegraphics[width=0.5\textwidth]{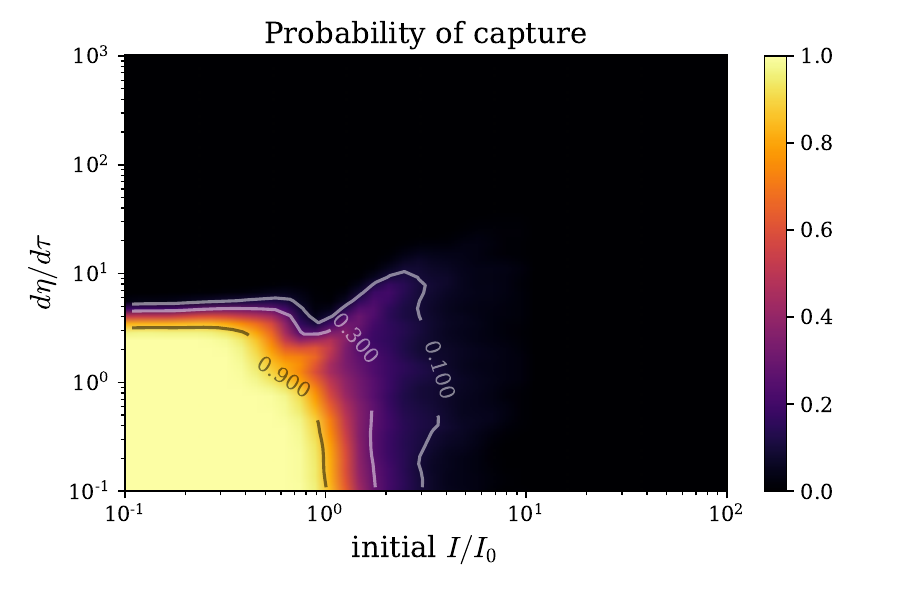}\includegraphics[width=0.5\textwidth]{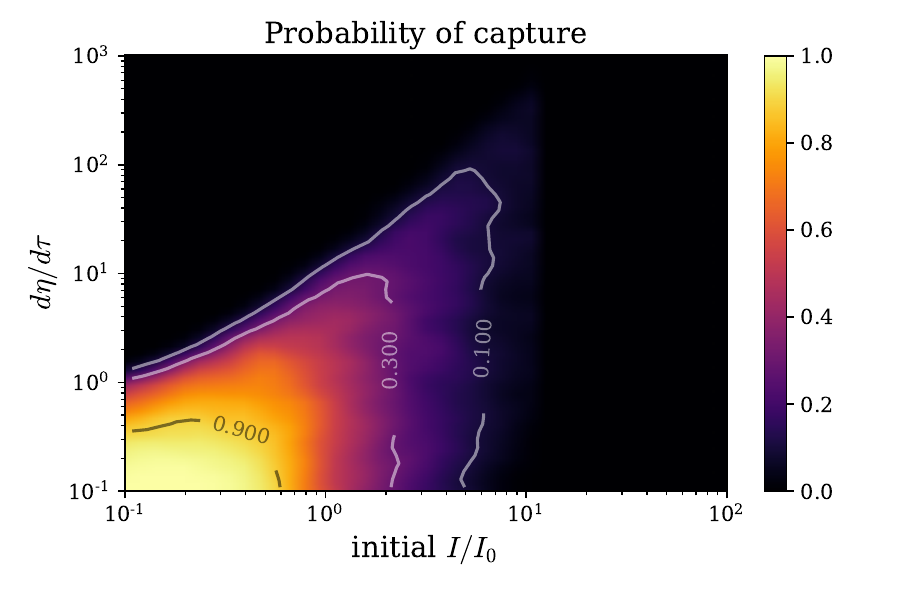}
\caption{Probability of capture as a function of the initial inclination $I$ (in unit of $I_0$) and the parameter $d\eta/d\tau$ (which is proportional to the migration rate).
\textbf{Left panel}: First-order ivection resonance and eccentric ivection resonance [given by the Hamiltonian \eqref{eq:Ham_1st}].
\textbf{Right panel}: Second-order ivection resonance [given by the Hamiltonian \eqref{eq:Ham_2nd}].
Contours of $10\%,30\%,90\%$ capture probability are plotted for reference.}
\label{capture_probability}
\end{figure}

In the absence of dissipative mechanisms, the maximum mutual inclination that initially coplanar planets can reach is $\sim I_0$, which is usually a few degrees or less, given its dependence on $\mu$ [see \eqref{eq:I0_1st}, \eqref{eq:I0_2nd} and \eqref{eq:I0_ecc}].
Moreover, the width of the resonance (i.e. the region in parameter space where the inclination of an initially coplanar system can be nontrivially excited) is small, making the probability of a system forming near an ivection resonance low.

However, planets often undergo migration after their formation, which leads to a smooth variation of $\dot\Omega$ that increases the likelihood for a system to encounter an ivection resonance during its migration. As we will show below, if the migration is in the desired direction and is sufficiently slow, the inclination of the system can be excited to $\gg I_0$ via resonance capture.

{In this section, we first study the process of ivection resonance capture and find the conditions of capture. This is mainly done with the help of the Hamiltonian model, which relates the ivection resonances to other previously studied resonances. Then we discuss the possible outcomes of ivection resonance, with a focus on the disruption of ivection resonance due to the encounter with another resonance, which cannot be captured by the Hamiltonian model.}

\subsection{{Ivection resonance capture}}
First we study the outcomes when the system encounters an ivection resonance during migration.
The dimensionless Hamiltonians \eqref{eq:Ham_1st} and \eqref{eq:Ham_2nd} have forms identical to first and second-order mean-motion resonances (MMR) up to some sign changes, so many of the results for MMR capture can be directly applied. The similarity can be easily seen from the contour of the Hamiltonians shown in Figure \ref{Ham_contour}. Due to this similarity, here we only summarize the possible outcomes of a resonance encounter. {(For of MMR capture and its outcomes, see \citealt{Peale76,BorderiesGoldreich84,MustillWyatt11,XuLai17}.)}

Near the resonance, $\eta$ is no longer constant since migration (or change in binary separation) changes $n_B$ or $\dot\Omega_0$. The speed and direction of migration can be characterized by the parameter $d\eta/d\tau$, where $\tau=t/T_0$ is the dimensionless time associated with the Hamiltonian. When the system encounters the resonance with $d\eta/d\tau<0$, it cannot be captured into the resonance. The inclination may increase by at most $\sim I_0$ as the system crosses the resonance.

When $d\eta/d\tau > 0$, resonance capture becomes possible.
Whether the system can be captured into the resonance depends on the migration rate and the initial mutual inclination.
When migration is slow ($d\eta/d\tau\lesssim 1$) and initial inclination small ($I/I_0\lesssim 1$), resonance capture is guaranteed. When migration is fast ($d\eta/d\tau\gg 1$) or initial inclination is large ($I/I_0\gg 1$), resonance capture is impossible. Between these two limits, capture is in general probabilistic. We numerically compute the probability of capture as a function of $d\eta/d\tau$ and initial $I/I_0$ {following the method in \citet{MustillWyatt11} (see their Figure 2, which presents the same result with slightly different notations)}, and the result is shown in Figure \ref{capture_probability}.

\begin{figure}
\centering
\includegraphics[width=.5\textwidth]{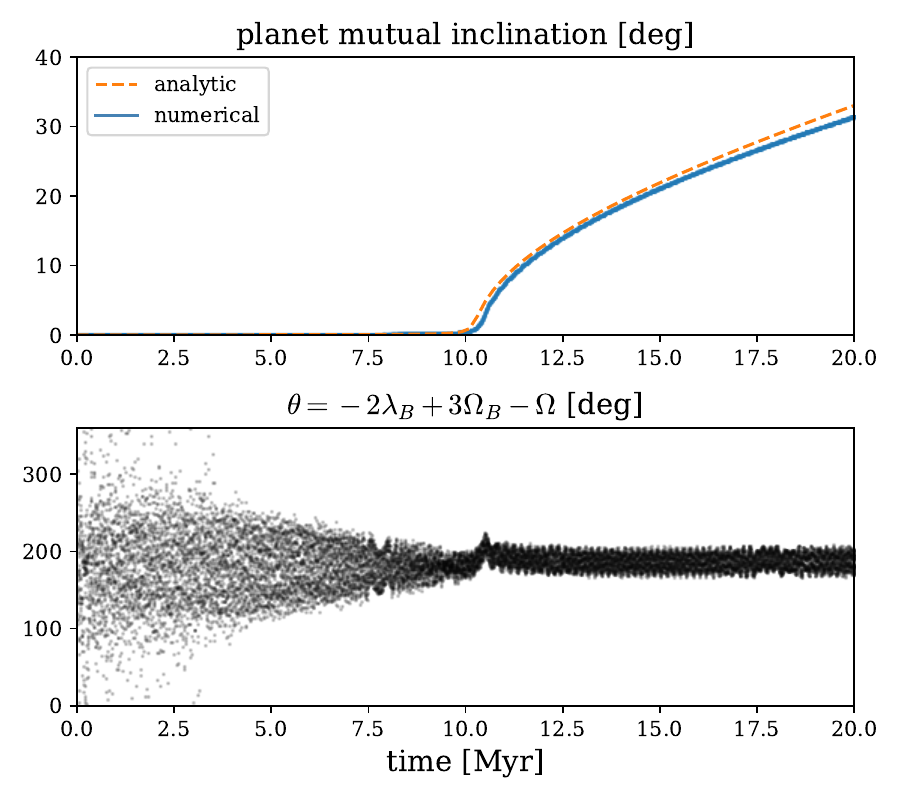}
\caption{An example of capturing into a first-order ivection resonance. The primary and the binary are both $1M_\odot$ stars, on circular orbit with $a_B\approx 200$ AU and $I_B=120^\circ$ initially. The two planets have mass $m_1=10M_\oplus$, $m_2=10M_{\rm Jup}$ and period $P_1=1$ year, $P_2=3.7$ year respectively, with initially circular and coplanar orbits. The binary migrates outward at a timescale of 40 Myr, increasing $|\dot\Omega|/n_B$. The system is captured into a first-order ivection resonance at $\sim$10 Myr. Once captured, the resonant angle librates around $\pi$ and the mutual inclination keeps increasing; meanwhile, the eccentricities of both planets remain small.
{The orange dashed curve in the top panel shows an analytic prediction of mutual inclination evolution based on the Hamiltonian \eqref{eq:Ham_1st}. Both the location of the resonance and the inclination evolution after resonance capture agree well with the numerical result up to $I \sim 30^\circ$.}}
\label{capture_outcome_0}
\end{figure}

An example of capturing into a first-order ivection resonance is shown in Figure \ref{capture_outcome_0}. For this example, the primary and the binary are both $1M_\odot$ stars, on circular orbit with $a_B\approx 200$ AU and $I_B=120^\circ$ initially. The two planets have mass $m_1=10M_\oplus$, $m_2=10M_{\rm Jup}$ and period $P_1=1$ year, $P_2=3.7$ year respectively, with initially circular and coplanar orbits. The system initially has {$n_B>\frac 12 |\dot\Omega|$}; to reach the resonance, 
we let the binary migrate outward at a timescale of 40 Myr. (Migrating the binary instead of the planets avoids crossing mean-motion resonances, the effect of which we discuss in Section \ref{subsec:disruption}.)
We integrate the system using the \textit{MERCURY} integrator \citep{Chambers12},
{with migration modeled with a user-defined force (see Eq. 75 in \citealt{XuLai17})}.
The system is captured into a first-order ivection resonance at $\sim$10 Myr. Once captured, the resonant angle librates around $\pi$ and the mutual inclination keeps increasing.

{We compare the numerical result with analytic prediction of the Hamiltonian \eqref{eq:Ham_1st} by plotting the inclination corresponding to the Hamiltonian's fixed point (around which a system captured into resonance should librate) in the top panel of Figure \ref{capture_outcome_0}.
This gives an analytic estimate for the evolution of mutual inclination.
This analytic estimate agrees well with the numerical result up to $I\sim 30^\circ$.
Note that we assumed small inclination when developing the Hamiltonian model, and this assumption naturally breaks down for large ($\gtrsim 1$ rad) mutual inclination.}

Systems captured into other types of ivection resonance show similar behavior (although the angle around which the resonant angle librates may be different).
In general,
if the system is captured into an ivection resonance, in the $(x,y)$ phase space it will librate around a stable fixed point of the Hamiltonian located at (for $\eta\gtrsim 1$)
\eq{
I/I_0 \approx \sqrt{x^2+y^2} \approx \sqrt{\eta/2}.
}
As $\eta$ continues to increase, $I^2$ increases approximately linear in time.
In reality, the growth of inclination stops when the migration rate changes such that $\eta$ no longer increases, or when the system is knocked out of resonance when passing another resonance (such as a MMR; see more in Section \ref{subsec:disruption}).
In either case, the final inclination is not directly limited by $I_0$, which is usually small for small $\mu$.

\subsection{Conditions of ivection resonance capture}
To capture the system into resonance, there are two main requirements: $\eta$ needs to be increasing when the system encounters resonance,
and the migration rate needs to be sufficiently slow ($d\eta/d\tau\lesssim 1$).\footnote{There is a third requirement that the initial mutual inclination should be $\lesssim I_0$. This condition is easily satisfied for typical giants with $\mu\gtrsim 10^{-3}$, since the initial inclination should be no more than a few degrees if the planets form in a mostly planar disk.}
Here we discuss the physical meaning of these requirements.

\subsubsection{Direction of migration}
The requirement that $d\eta/d\tau >0$
physically means that the ratio between the precession rate and the binary frequency $|\dot\Omega_0|/n_B$ needs to be increasing.
If we assume that $n_B$ remains fixed, this requires the planets to migrate convergently (with increasing $a_1/a_2$) or inward.\footnote{Having convergently \textit{or} inward migrating planets is a necessary but insufficient condition for having an increasing $|\dot\Omega_0|$.}

It is worth noting that the direction of migration required for evection resonance capture is usually the opposite. Evection resonance capture requires the apsidal precession rate $|\dot\varpi|$ to be decreasing, which requires the planets to migrate outward or divergently.

\subsubsection{Critical migration timescale}\label{subsec:Tcrit}
Ivection resonance capture also requires the migration to be sufficiently slow. The physical timescale of migration can be characterized by the timescale of the evolution of $|\dot\Omega_0|$,
\eq{
T_\Omega\equiv \left|\frac{d\ln|\dot\Omega_0|}{dt}\right|^{-1}.
}
Normally, this timescale is comparable to the minimum of the planets' migration timescale (defined as $T_{m,i}\equiv|\dot a_i/a_i|^{-1}$).

$T_\Omega$ is related to the dimensionless parameter $|d\eta/d\tau|$ via (assuming fixed $n_B$)
\eq{
\left|\frac{d\eta}{d\tau}\right| = \frac 12 T_0^2|\dot\Omega_0| T_\Omega^{-1}.
}
Therefore, the requirement $|d\eta/d\tau|\lesssim 1$ corresponds to
\eq{
\min(T_{m,1},T_{m,2}) \sim T_\Omega \gtrsim T_{\rm crit},\label{eq:Tm_req}
}
where the critical migration timescale $T_{\rm crit}$ is defined as
\eq{
T_{\rm crit} \equiv \frac 12 T_0^2 |\dot\Omega_0|.
}
Note that $\min(T_{m,1},T_{m,2})\gtrsim T_{\rm crit}$ means that the time it takes for the system to migrates across the width of the resonance (i.e. changing $\eta$ by a few) is longer than the timescale of libration. 

The requirement \eqref{eq:Tm_req} is a relatively coarse estimation, mainly because the maximum $d\eta/d\tau$ for which capture probability is nontrivial and can vary by a couple orders of magnitude depending on the initial inclination of the system. For instance, the probability of capture is still $\sim 10\%$ when $d\eta/d\tau\sim 10$ for first-order ivection resonance and eccentric ivection resonance and $\sim 100$ for second-order ivection resonance, \emph{if} the initial inclination is optimal (see Figure \ref{capture_probability}).

In the limit of small $a_1/a_2$, $T_{\rm crit}$ is given by
\eq{
T_{\rm crit} \approx \left\{
\begin{array}{ll}
146{\rm ~Myr}\times \left(\frac{P_2}{1{\rm ~yr}}\right)\left(\frac{P_1}{P_2}\right)^{-28/9}\left(\frac{\mu}{10^{-3}}\right)^{-7/3}\left[\frac{q}{1+q}(1-\cos I_B)|\sin I_B|\right]^{-4/3} & \text{1st order}\\
5.4{\rm ~Gyr}\times \left(\frac{P_2}{1{\rm ~yr}}\right)\left(\frac{P_1}{P_2}\right)^{-6}\left(\frac{\mu}{10^{-3}}\right)^{-3}
\left(\frac{q}{1+q}\right)^{-2}(1-\cos I_B)^{-4}& \text{2nd order}\\
0.23{\rm ~Myr}\times \left(\frac{P_2}{1{\rm ~yr}}\right)\left(\frac{P_1}{P_2}\right)^{-28/9}\left(\frac{\mu}{10^{-3}}\right)^{-7/3}
\left(\frac{1-e_B}{0.05}\right)^2j^{8/3}\left(\frac{1+q}{q}|\sin 2I_B|\right)^{-4/3}& \text{eccentric}
\end{array}
\right.
\label{eq:Tcrit_est}
}
This can still be used as a coarse estimation of $T_{\rm cirt}$ when $a_1$ and $a_2$ are comparable. For eccentric ivection resonance, the integer $j$ is the ratio between planet precession rate and binary orbital frequency ($|\dot\Omega_0|\approx jn_B$).
From this estimation, we can see that $T_{\rm crit}$ tends to be relatively large for ivection resonance with a circular binary.
Especially, the large $T_{\rm cirt}$ makes capturing into a second-order ivection resonance very unlikely, unless the planets are very massive (e.g. $\mu\gtrsim 10^{-2}$).
Meanwhile, if the binary is highly eccentric (e.g. with $e_B>0.9$ or $0.95$), $T_{\rm crit}$ can easily becomes smaller than the typical planet migration timescale.
Equation~\eqref{eq:Tcrit_est} also shows that $T_{\rm crit}$ has a strong dependence on the planet mass (the $\mu$ parameter). Therefore, among systems that have binary companions with suitable period, 
ivection resonance is significantly more likely to affect those with more massive planets,

\subsection{\red{Possible outcomes of ivection resonance capture}}\label{subsec:disruption}
So far we mainly discuss the process of ivection resonance capture, which can be well described by the Hamiltonian model. The final outcome of ivection resonance capture, i.e. the final state of the system that we observe today, depends heavily on other factors, such as the encounter with a different resonance. In this section, we broadly discuss the possible outcomes of ivection resonance capture.

\subsubsection{Maintaining ivection resonance}
If the planets have not migrated very far after they are captured into the ivection resonance, so that they never encounter a second resonance or reach very large mutual inclination, they will likely remain inside the ivection resonance.
This usually leads to a modestly large mutual inclination, small eccentricities (unless eccentricities have been excited before encountering the ivection resonance), and the resonant angle should librate around $\pi$ for first-order ivection resonance or eccentric ivection resonance with prograde binary, $0$ for eccentric ivection resonance with retrograde binary, and $\pm\pi/2$ for second-order ivection resonance.

(or $0$ if the resonance is an eccentric ivection resonance and the binary is retrograde).
The mutual inclination today should be of order
\eq{
I \approx I_0\sqrt{\eta} \sim \left(1-\frac{\partial\theta}{\partial\lambda_B}\frac{n_B}{|\dot\Omega_0|}\right).
}
This gives $I = \mathcal O(1)$ rad if migration changes $n_B/|\dot\Omega_0|$ by $\mathcal O(1)$ after the resonance capture.

\subsubsection{Disruption of ivection resonance}
In reality, a migrating system is very likely to encounter another (secular or mean-motion) resonance before migration stops. In most (if not all) scenarios, encountering another resonance will disrupt the ivection resonance.

\begin{figure}
\centering
\includegraphics[width=\textwidth]{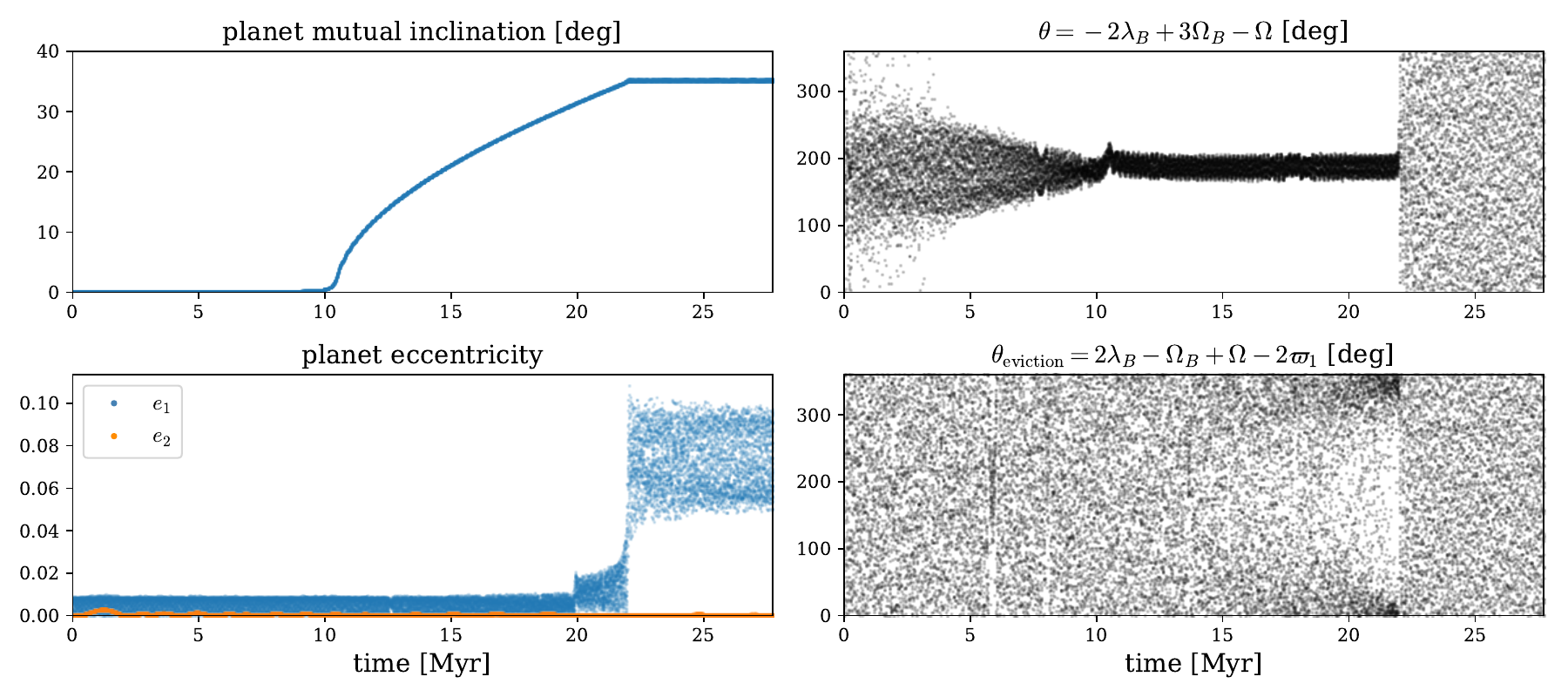}
\caption{Disruption of ivection resonance due to crossing an eviction resonance, when extending the simulation shown in Figure \ref{capture_outcome_0}. The top right and bottom right panel shows the resonant angle of ivection resonance and eviction resonance respectively. After encountering the eviction resonance, the mutual inclination ceases to increase.}
\label{capture_outcome_1}
\end{figure}

For example, if we integrate the system in Figure \ref{capture_outcome_0} for longer time, the system will eventually encounter an eviction resonance, which is another secular-orbital resonance with resonant interaction $\propto e_i^2 I$ and resonant angle $2\lambda_B - 2\Omega_B+\Omega-2\varpi_i$ \citep{ToumaWisdom98}. The system's evolution is shown in Figure \ref{capture_outcome_1}. When the system gets close to the eviction resonance, it is affected by both ivection and eviction resonance, making the dynamics of the system chaotic while the two resonances overlap.
This chaotic behavior knocks the system out of the ivection resonance's resonant zone (i.e. the region in the phase space in which the resonant angle librates).
The inclination ceases to increase, and the planets end up being out of resonance, with nearly constant inclination and small eccentricity (excited during the transient chaos).
This is the typical outcome of encountering a relatively weak resonance whose with is narrower and the libration timescale is longer then the ivection resonance (but still short compared to the migration timescale; otherwise the effect of that resonance will be totally negligible).
In principle, it is possible for the system to remain inside ivection resonance after encountering a relatively weak resonance, but it should be very rare (because it requires the system to luckily return to the resonant zone after the transient chaos) and we never observe it in our simulations.

\begin{figure}
\centering
\includegraphics[width=\textwidth]{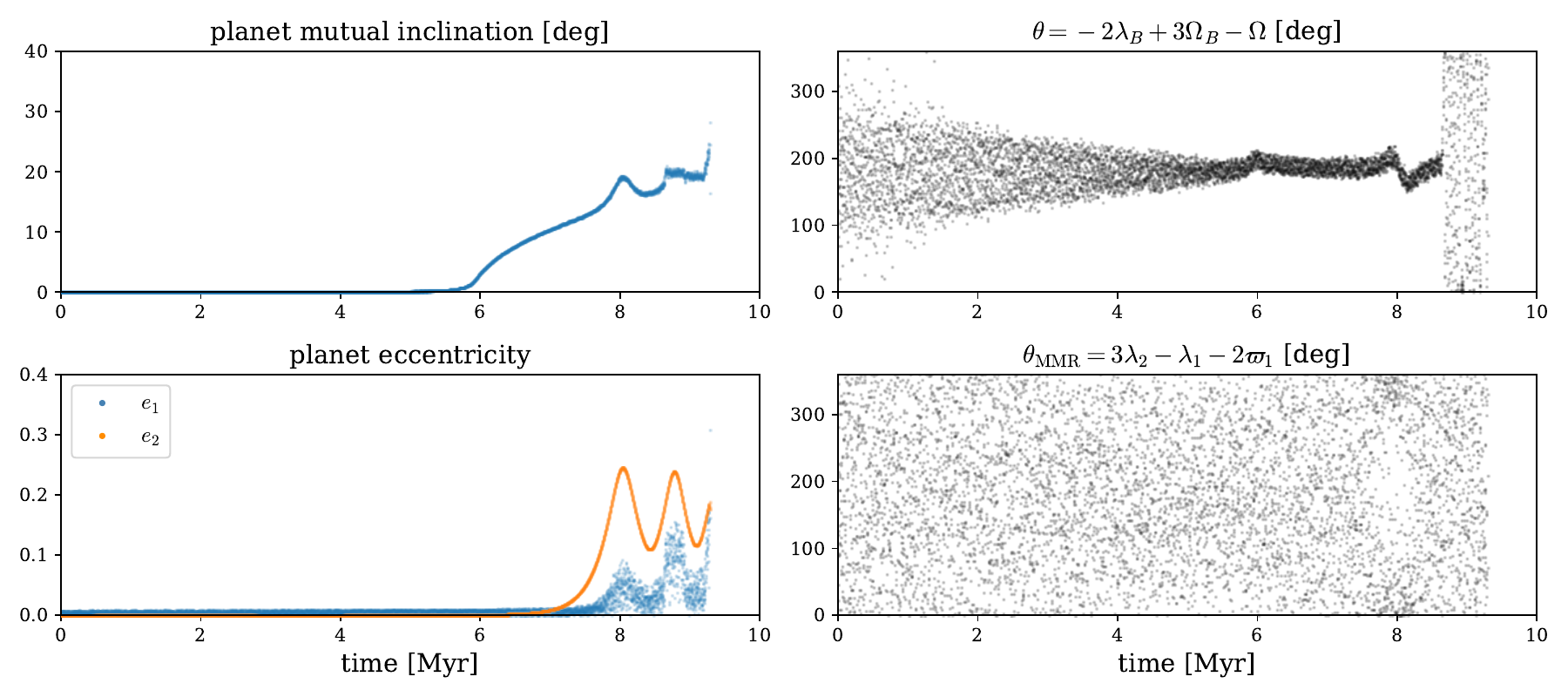}
\caption{Disruption of ivection resonance due to crossing a MMR.
The setup is similar to that of Figure \ref{capture_outcome_0} and \ref{capture_outcome_1}, except that the binary orbit is fixed and the outer planet migrates inward.
After encountering the 1:3 MMR, the eccentricities of both planets increas significantly, and the inner planet is eventually ejected {at 9.3 Myr.}}
\label{capture_outcome_2}
\end{figure}

Another example is when the system encounters a relatively strong MMR after being captured into an ivection resonance.
This is relevant for real systems where one or both planets migrate and their semi-major axis ratio changes.
For instance, consider a system with the same parameters as that shown in Figure \ref{capture_outcome_1}, except that the binary has fixed orbit at 215 AU and the outer planet migrates inward with a timescale of 80 Myr.
The system encounters a 1:3 MMR after being captured into an ivection resonance.
As the system approaches 1:3 MMR, the eccentricities of both planets are excited.\footnote{This happens before the system actually reaches 1:3 MMR, because MMR is a much stronger resonance and its perturbation already becomes significant compared to that of ivection resonance when the resonant angle for MMR is still circulating. As a side note, the outer planet, which is much more massive than the inner planet, attains larger eccentricity than the inner planet; the reason of that is still unclear.}
Eventually, the system is forced out of the ivection resonance, and the inner planet becomes unstable and is ejected soon after that.
This is the typical outcome of encountering a resonance stronger than the ivection resonance: The ivection resonance will be disrupted, and if the inclination is sufficiently large when the system reaches the second resonance, the smaller one of the two planets may be ejected.

It is worth noting that the relative strength of the encountered resonance can depend on the inclination at which the encounter happens. For instance, for the inclination mode of a second-order MMR, the resonant forcing in inclination is $\propto I$, making it stronger than ivection resonance if encountered at sufficiently large $I$ but weaker if encountered at very small $I$.

\subsubsection{Other possibilities}
There are a few other possible outcomes that we have not discussed above; these scenarios are either less common, or involves physics beyond the scope of this paper. One possibility is that the system manages to migrate a long way inside ivection resonance (without being disrupted) and reach large inclination. In this scenario, the evolution has to be studied with numerical simulations since the Hamiltonian model no longer applies. We do not consider this possibility in this paper because in reality the system will most likely encounter another resonance - either another secular resonance or a MMR - before this happens.

Another important mechanism that we have not discussed is the damping of inclination. When strong enough, this can prevent the mutual inclination from reaching large values or significantly damp the mutual inclination after the system leaves ivection resonance. Moreover, inclination damping may change the stability of resonant libration. Without inclination damping, libration is stable; but including damping may make libration overstable (i.e. the amplitude of $\theta$ libration increases for every cycle), which eventually drives the system out of resonance. This mechanism is essentially the same as the resonance escape mechanism for MMR discussed in \citet{GoldreichSchlichting2014}.
The effect of inclination damping to the outcome of ivection resonance capture should be studied in more details in future studies.

\section{Formation of Kepler-108}\label{sec:K108}

\begin{figure}
\centering
\includegraphics[width=\textwidth]{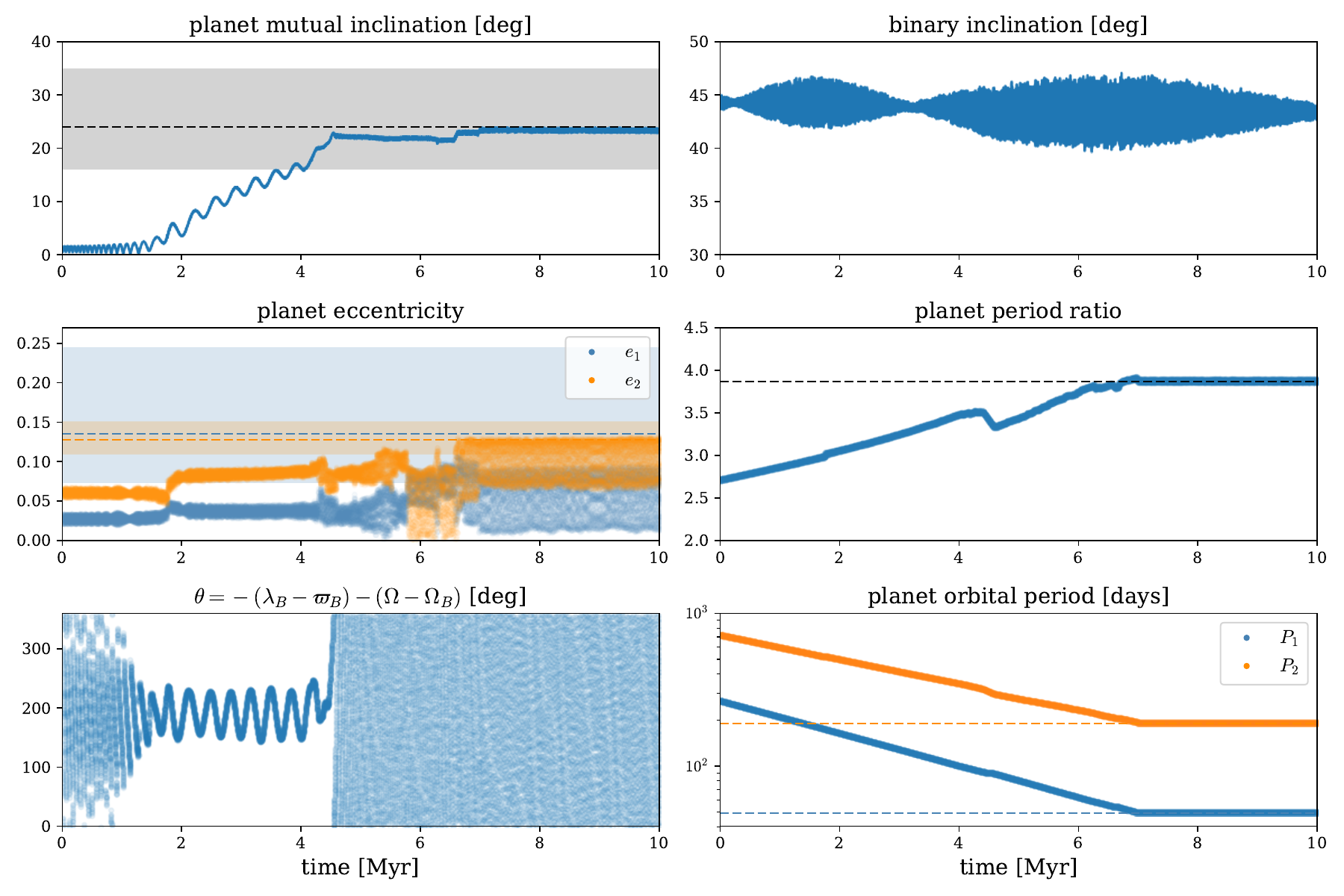}
\caption{A simulation that reproduces the orbital configuration of Kepler-108 from initially coplanar planets. Dashed lines in the top left, middle left, middle right and bottom right panels show observed system parameters, with the 1-$\sigma$ observational uncertainty of eccentricity (middle left) and inclination (top left) marked by shades.
The mutual inclination is excited due to capturing into an eccentric ivection resonance with a $e_B=0.95,I_B=45^\circ, P_B=7000$ yr binary.
The system leaves the resonance (at $t\sim 4.5$ Myr) probably due to encountering in a 2:7 MMR.
The initial eccentricities are due to a passage through 1:2 MMR at $t<0$.
The eccentricities are further excited when passing a 1:3 MMR at $\approx 1.8$ Myr and after the ivection resonance is disrupted by the 2:7 MMR.
The final eccentricities and mutual inclination are consistent with observation.
See Section \ref{sec:K108} for more discussion of this simulation.}
\label{K108}
\end{figure}

In this section we apply the results from previous sections to discuss a possible formation scenario of Kepler-108 \citep{MillsFabrycky17}.
The system hosts two planets with eccentricity $e_i\sim 0.1$ and mutual inclination $I\approx 24^\circ$,
and is perturbed by a binary companion with a sky projected separation of $\sim 300$ AU.
The orbital period of the binary is comparable to the timescale of nodal precession (which is $\approx 5700$ years), suggesting that ivection resonance may have played an important role in the formation of the system.\footnote{
Mutual inclination can also be excited when another planet is scattered out of the system, or by Lidov-Kozai oscillation.
However, exciting the mutual inclination of Kepler-108 through these mechanisms are unlikely, since
exciting such significant mutual inclination by planet-planet scattering tends to produce larger eccentricities than the values observed in Kepler-108, and Lidov-Kozai mechanism is completely suppressed by the fast precession of the planets (see Section 6 of \citealt{MillsFabrycky17}).}

To model the migration of the planets, we assume that the planets start far from the resonance and migrate with constant $T_{m,i}\equiv |\dot a_i/a_i|^{-1}$ for a given amount of time, then the migration stops and we wait until the system reaches a steady state.
This model is oversimplified in that it does not capture the time-dependence of the migration rate, the precession of the planets caused by the disk, and possible eccentricity and inclination damping. Still, we expect that these oversimplifications do not qualitatively affect the results.

{We attempt to reproduce the currently observed orbital configuration using an eccentric ivection resonance, because the resonant perturbation will be much weaker and resonance capture will require unrealistically slow migration if the binary has small eccentricity instead.
We fix the binary eccentricity at $0.95$ and initial inclination $I_B=45^\circ$. These parameters only affect the strength of the resonance; for a smaller binary eccentricity, we can get similar results by correspondingly decreasing the migration rates.
We assume that both planets migrate inwards, and fix $T_{m,2}$ at 8 Myr. The initial semi-major axes and the time to stop migration are chosen to reproduce the observed planet semi-major axes, and we tune $T_{m,1}$ and the binary period $P_B$ (which together determine when eccentric ivection resonance and MMRs are encountered) to reproduce the observed eccentricity and inclination.}

Figure \ref{K108} shows an example that manages to reproduce the currently observed orbital configuration of the system. The planet and stellar masses are set to the observed values.
For this simulation, $T_{m,1}=6.2$ Myr and $P_B=7000$ yr. In this case the two planets migrate divergently (so they cannot be captured into any MMR) but the nodal precession rate $|\dot\Omega|$ increases.
{We start the simulation at $t=-7$ Myr, with a planet period ratio $<2$.
After passing a 2:1 MMR at $t\sim -5.5$ Myr, the planets gain some finite eccentricities $e_1\sim 0.03,~e_2\sim 0.06$.
(The evolution for $t<0$ is not shown in Figure \ref{K108} since we want to focus on the evolution of the system after it gets close to the ivection resonance.)
Having some finite eccentricity before the system encounters ivection resonance is necessary in order to produce eccentricities consistent with observation (because ivection resonance does not excite eccentricity), but such eccentricity need not come from crossing a 1:2 MMR.}

At $t\sim 1$ Myr, the system approaches and gets captured into an eccentric ivection resonance (with $|\dot\Omega|\approx n_B $) and the resonant angle begins to librate.
The finite ``initial" eccentricity does not significantly affect capturing into the ivection resonance, because there is no low-order coupling between eccentricity and inclination.
Once the system is inside the resonance, inclination begins to increase as migration drives the system deeper into resonance.
{Shortly after the capture, the system crosses 1:3 MMR at $t\sim 2$ Myr, which further increases the planet eccentricities.
Crossing this 1:3 MMR does not disrupt the ivection resonance, because the mutual inclination and planet eccentricities are all still small, and the forcing due to a 1:3 MMR - which is $\propto I$ for the inclination mode and $\propto e_i$ for the eccentricity mode - is weak.}

The ivection resonance is disrupted at $T\sim 4.5$ Myr, and the resonant angle ceases to librate. This is likely due to the encounter of a 2:7 (1:3.5) MMR, which allows the coupling between eccentricity, inclination and semi-major axis. Note that since the mutual inclination of the system is large, the $I^4 e$ mode of this resonance has strength comparable to a first-order MMR.
Once the system is no longer inside ivection resonance, the mutual inclination stops increasing. Secular coupling between the planets cause the eccentricities to oscillate with relatively large amplitude.

We stop the migration at $t=7$ Myr, when the semi-major axes of the planets reach the observed values. The system ends up in a quasi-steady state where the inclination is nearly constant and the eccentricities oscillate at constant amplitude. The final eccentricities are slightly smaller than the observed value, but the difference is within 1-$\sigma$ of observational uncertainty.


{This example does require a certain amount of fine-tuning: We tuned the relative locations of the ivection resonance and MMRs so that the 2:7 MMR can disrupt the ivection resonance at desired inclination and the 1:3 MMR is encountered while $I$ is still small (or before encountering the ivection resonance), avoiding it to disrupt the ivection resonance.
Still, the qualitative behavior that the system ends up in a non-resonant state with significant $I$ and smaller $e_i$ is fairly generic, and can be observed for a wide range of parameter choice (as long as ivection resonance capture is allowed)
The main limiting factor that may prevent such behavior is that capturing into ivection resonance requires large binary eccentricity or very slow migration. As we discuss in the next section, this is also why ivection resonance capture may be relatively uncommon.
Overall, we do not intend to claim that this example, with an oversimplified migration model, represents the actual formation scenario of Kepler-108. (For example, the actual eccentricity evolution is likely different.)
Instead, it illustrates that ivection resonance is a promising mechanism for exciting inclinations in systems like Kepler-108 which have significant mutual inclination and modest eccentricities today.}

\begin{longrotatetable}
\begin{deluxetable*}{lcccccc}
\tablewidth{700pt}
\tablecaption{The scaling of parameters of ivection and evection resonances\label{scalings}}
\tablecolumns{7}
\tablehead{
\colhead{Name}&
\colhead{Order}&
\colhead{Resonant angle}&
\colhead{$e_0$ or $I_0$}&
\colhead{$T_0$}&
\colhead{$T_{\rm crit}$}&
\colhead{Capture requirement}
}
\startdata
1st order ivection resonance & 1 & $-2\lambda_B+3\Omega_B-\Omega$ & $\mu^{1/3}$ & $\mu^{-2/3}P_B$ & $\mu^{-4/3}P_B$ & increasing $|\dot\Omega|/n_B$\\
2nd order ivection resonance & 2 & $-\lambda_B+2\Omega_B-\Omega$ & $\mu^{1/2}$ & $\mu^{-1}P_B$ & $\mu^{-2}P_B$ & increasing $|\dot\Omega|/n_B$\\
eccentric ivection resonance & 1 & $-j(\lambda_B-\varpi_B)-(\Omega-\Omega_B)$ & $\mu^{1/3}(1-e_B)^{-1/2}$ & $\mu^{-2/3}(1-e_B)P_B$ & $\mu^{-4/3}(1-e_B)^2P_B$ & increasing $|\dot\Omega|/n_B$\\
\hline
evection resonance & 2 & $\lambda_B-\varpi_i$ & $\mu^{1/2}$ & $\mu^{-1}P_B$ & $\mu^{-2}P_B$ & decreasing $|\dot\varpi_i|/n_B$\\
\hline
1st order MMR & 1 & $(j+1)\lambda_2-j\lambda_1-\varpi_i$ & $\mu^{1/3}$ & $\mu^{-2/3}P$ & $\mu^{-4/3}P$ & increasing $n_1/n_2$\\
2nd order MMR & 2 & $\frac 12(j+2)\lambda_2-\frac 12j\lambda_1-\varpi_i$ & $\mu^{1/2}$ & $\mu^{-1}P$ & $\mu^{-2}P$ & increasing $n_1/n_2$\\
\enddata
\tablecomments{Parameters for MMR are also given for reference. We assume the planet periods satisfy $P_1\sim P_2\sim P$. For ivection and evection resonances, the planet and binary period needs to satisfy $P_B\sim \mu^{-1}P$. $j$ can be any (relatively small) positive integer. $e_0$ or $I_0$ physically corresponds to the maximum eccentricity (for evection resonance and MMR) or inclination (for ivection resonances) the system can gain when passing a resonance without capture. $T_0$ gives the timescale of libration, and $T_{\rm crit}$ gives the minimum migration timescale required for capture.}
\end{deluxetable*}
\end{longrotatetable}

\section{Application to other exoplanet systems}\label{sec:application}
\begin{figure}
\centering
\includegraphics[width=.6\textwidth]{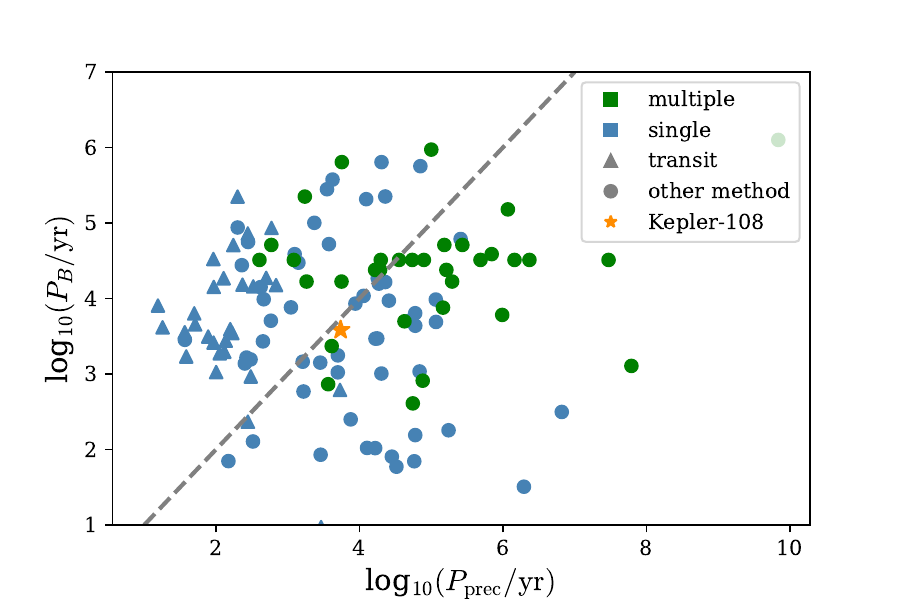}
\caption{Estimates for the nodal precession timescale $P_{\rm prec}$ and binary period $P_B$ of systems with known external binary companion. Color marks whether the system hosts multiple observed planets, and marker shape corresponds to detection method. The grey dashed line marks $P_{\rm prec}=P_B$.}
\label{PB_vs_Pprec}
\end{figure}

\begin{figure}
\includegraphics[width=.5\textwidth]{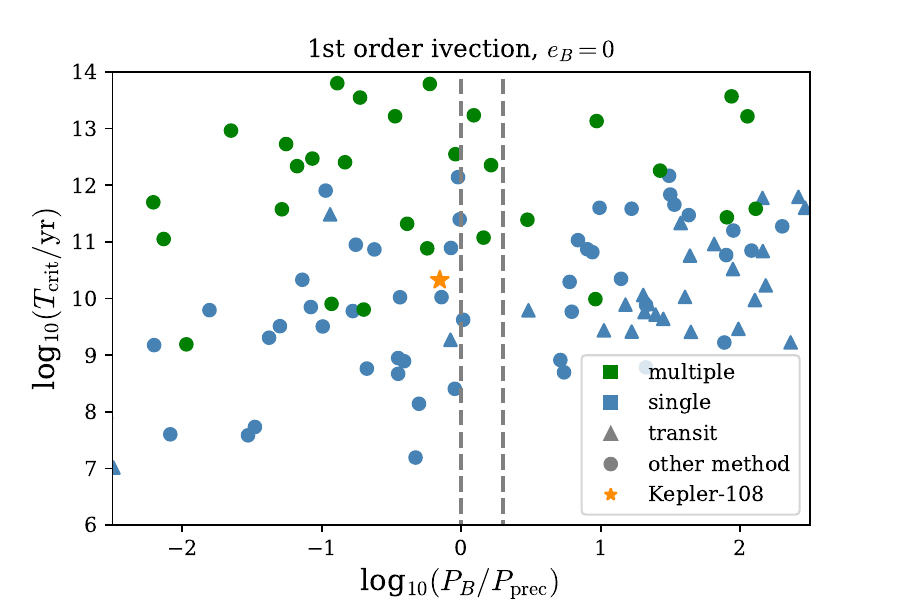}\includegraphics[width=.5\textwidth]{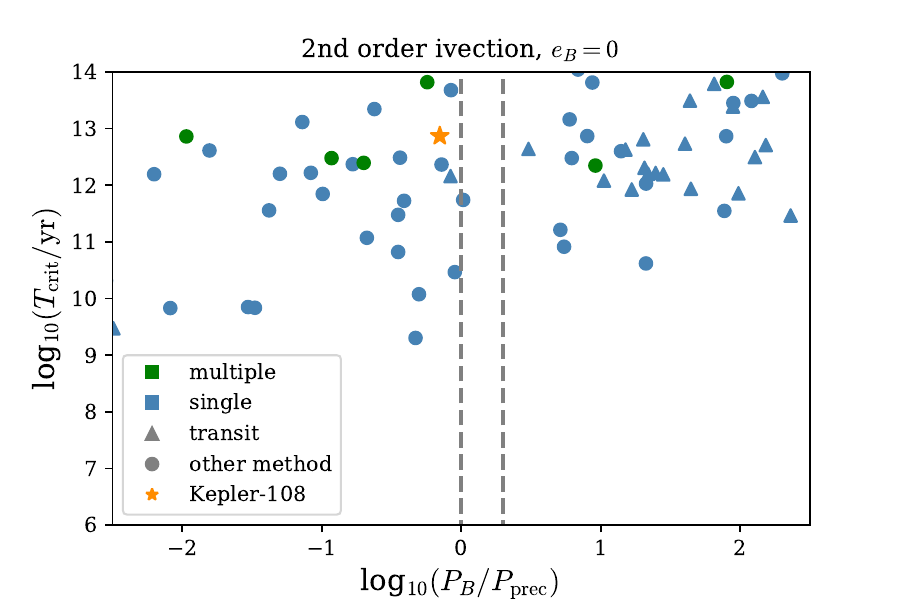}
\includegraphics[width=.5\textwidth]{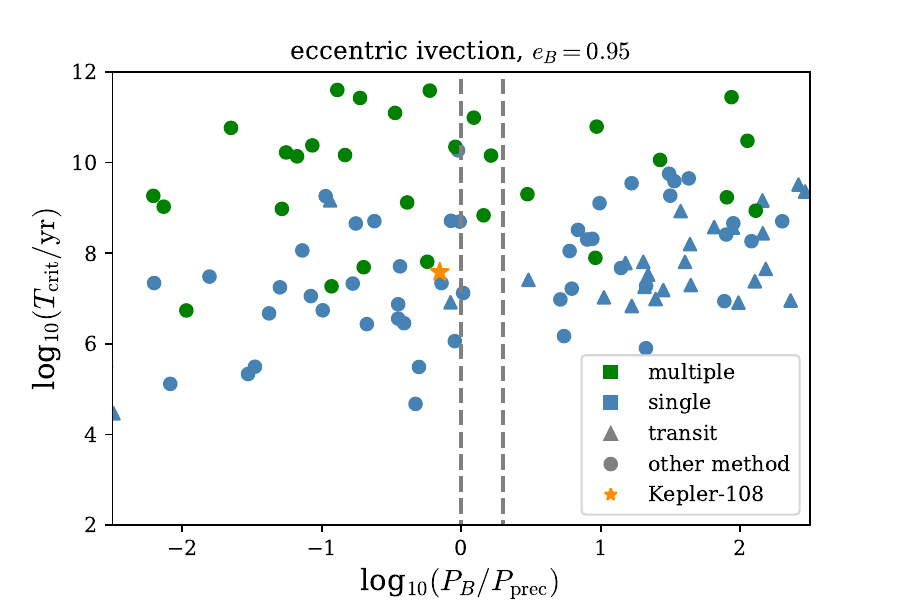}
\caption{Estimates of the critical migration timescale $T_{\rm crit}$ and the ratio between $T_{\rm prec}$ and $P_B$, evaluated for 1st order ivection resonance (top left panel; $e_B=0$), 2nd order ivection resonance (top right panel; $e_B=0$) and eccentric ivection resonance (bottom left panel; $e_B=0.95$). The grey dashed lines mark $P_B/P_{\rm prec}=1$ and 2.}
\label{T_crit_plt}
\end{figure}

The most direct consequence of ivection resonance capture is the excitation of mutual inclination.
However, even if exciting mutual inclination by ivection resonance is common, finding another system like Kepler-108 will still be very difficult since we seldom manage to observe the mutual inclination of planets (except when they are nearly coplanar).
In this section, we determine whether ivection resonance can be common among exoplanet systems with external binary companion using indirect evidences such as the overall statistics of the precession period and critical migration timescale.

\subsection{Can precession period match binary period?}
Ivection resonance happens when the period of nodal precession ($P_{\rm prec}$) is commensurate with the binary period ($P_B$). Therefore, we can infer the likelihood of passing ivection resonance during migration from the distribution of $P_{\rm prec}$ versus $P_B$.

This distribution of $P_{\rm prec}$ and $P_B$ is shown in Figure \ref{PB_vs_Pprec}. The precession period is estimated as follows: For single-planet systems, the precession rate is estimated by setting $\mu$ to the planet-star mass ratio and evaluate $f_3(\alpha)$ at $\alpha=2.57$, which is the median $\alpha$ for adjacent planets in multiplanet systems detected by RV.\footnote{Here we choose an $\alpha$ value representative of systems discovered by RV (instead of all systems, which would give median $\alpha=1.67$), because planets prone to ivection and evection resonances are mainly detected by RV due to their relatively large mass and semi-major axis.}
Physically, this gives the precession rate if the system hosts, or once hosted, another planet with similar or smaller mass.
For multiple-planet systems, the precession period of each pair of planets are evaluated. This overestimates the precession period, if there are more than two planets.

In Figure \ref{PB_vs_Pprec}, there is no significant correlation between $P_B$ and $P_{\rm prec}$, and many systems have $P_B\sim P_{\rm prec}$. More quantitatively, $\sim 10\%$ of the systems in Figure \ref{PB_vs_Pprec} can pass an ivection resonance if $T_{\rm prec}$ changes by one order of magnitude during migration. Note that the actual distribution may be different from Fig \ref{PB_vs_Pprec} since we do not account for observational bias and only include systems in which the mass (or $M\sin i$) of at least one planet is known in this figure.

\subsection{Is migration slow enough?}
Ivection resonance capture is also limited by the migration rate of the planets, with resonance capture possible only when migration timescale is no shorter than $T_{\rm crit}$.
The critical migration timescale $T_{\rm crit}$ corresponding to the systems in Figure \ref{PB_vs_Pprec} are shown in Figures \ref{T_crit_plt}. We estimate $T_{\rm crit}$ for first and second order ivection resonance at $e_B=0$ and for eccentric ivection resonance at $e_B=0.95$ using \eqref{eq:Tcrit_est}, assuming optimal $I_B$. When estimating $T_{\rm crit}$, we scale the semi-major axes of the planets (while maintaining the semi-major axis ratio) so that the nodal precession period is commensurate with the binary period; this partially accounts for the migration after the system passes the resonance.
For systems with a single observed planet, we use the approximation \eqref{eq:Tcrit_est} with $P_1/P_2=1$.

The first two panels of Figure \ref{T_crit_plt} show that when the binary eccentricity is small, $T_{\rm crit}$ tends to be large. 
Very few systems have relatively small $T_{\rm crit}$ (i.e. $\lesssim 10$ Myr) for first-order ivection resonance, and $T_{\rm crit}$ for second-order ivection resonance is at least $\sim 1$ Gyr.
Meanwhile, as shown in the last panel of Figure \ref{T_crit_plt}, when the binary eccentricity is large, it becomes much easier to have low $T_{\rm crit}$.
Therefore, we expect ivection resonance to be important only when binary eccentricity is large.

\begin{figure}
\centering
\includegraphics[width=.5\textwidth]{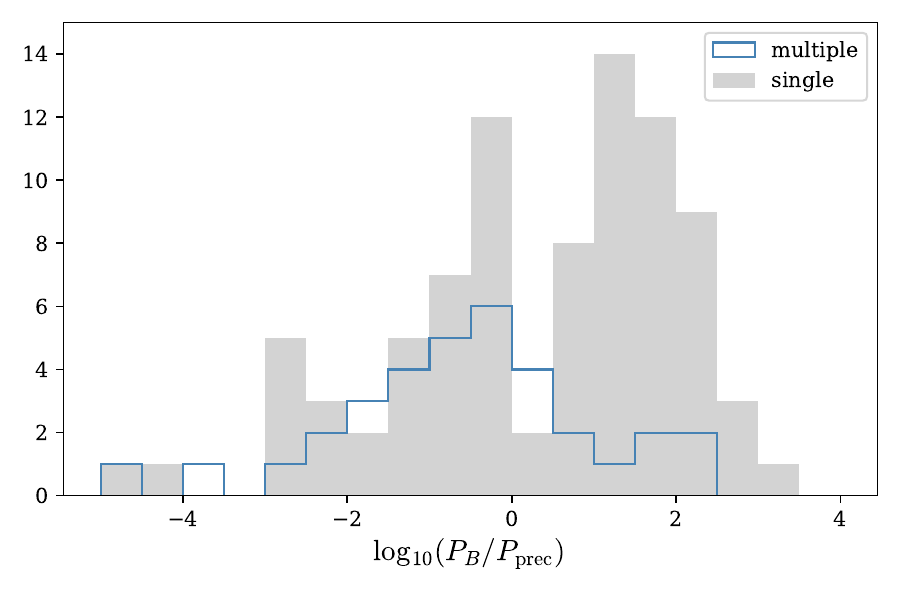}
\caption{The distribution of (estimated) $P_B/P_{\rm prec}$ for systems with a single observed planet and with multiple observed planets. A gap appears for single-planet systems at 0 - 0.5 dex (see analysis of the statistical significance in Appendix \ref{a:test}). If the gap is not a coincidence due to the small sample size, this will suggest that ivection / evection resonance may play an important role in the formation and evolution of exoplanet systems with binary companion.}
\label{Prat_distribution}
\end{figure}

\subsection{{An observational signature?}}\label{subsec:importance}
Although the critical migration timescale $T_{\rm crit}$ tends to be very large, ivection (and evection) resonance may still play an important role in the evolution of exoplanet systems with binary companion.
This can be noticed by inspecting the distribution of $P_B/P_{\rm prec}$ (Figure \ref{Prat_distribution}), which appears to show a gap at estimated $P_B/P_{\rm prec}$ between 0 and 0.5 dex for systems with a single observed planet.
{Meanwhile, such gap does not show up for systems with multiple observed planets, possibly because ivection (and evection) resonances are generally weaker in this sample (see the systematic difference in $T_{\rm cirt}$ between the samples with single and multiple observed planets in Figure \ref{T_crit_plt}).
A deficiency of systems close to ivection resonance is also visible in Figure \ref{PB_vs_Pprec}.}

{Given the relatively small sample size,
one might suspect that the gap in Figure \ref{Prat_distribution} is just a coincidence or the result of a particular binning. We show in Appendix \ref{a:test} that the gap is not an artifact due to binning, and the probability of observing a similar gap from a gap-less distribution is $\sim 6\%$.
We also comment that our estimates of $P_B$ and $P_{\rm prec}$ both have relatively large uncertainties (from estimating $a_B$ from sky-projected distance, using $M\sin i$ for planet mass, etc.), especially for systems with one observed planet (where the location of another hidden / ejected planet also comes from a coarse estimate). 
These uncertainties could smear out features in the distribution or slightly shift the location of features (via systematic errors in $P_B$ and $P_{\rm prec}$ estimates). However, they could not create a gap from a gap-less distribution as long as the errors are not strongly correlated with $P_B/P_{\rm prec}$. Therefore, the existence of a gap in the estimated distribution of $P_B/P_{\rm prec}$ should suggest that the true distribution of $P_B/P_{\rm prec}$ likely contains a more prominent gap.}


Given that the distribution of $P_B$ and $P_{\rm prec}$ are both wide and gap-less, the most reasonable explanation for the gap, if it is not a coincidence, is that it is associated with some mechanism that only operates at $P_B/P_{\rm prec}\sim 1$, which is most likely ivection or evection resonance.
One possibility is that
the gap is due to scattering of planets when a resonant system becomes dynamically unstable (e.g. due to encountering another strong resonance, as discussed in Section \ref{subsec:disruption}). This could either push the semi-major axes away from the resonant values, or cause ejection of the less massive planet.
If such mechanism has a strong effect, there should be a deficiency of planets in regions of the parameter space where it is more likely to be in resonance (i.e. when estimated $P_B/P_{\rm prec}\sim 1$).
If this is the case, the gap should exist in the distribution of both single and multi-planet systems, although it may be less observable for multi-planet systems due to the much smaller sample size.
{However, such dynamical instability requires the planets to be captured into an ivection (or evection) resonance in the first place.
As shown in Figure \ref{T_crit_plt}, for most systems the critical migration timescale for ivection resonance is extremely long, making resonance capture unlikely. 
Evection resonance suffer from the same problem, since
$T_{\rm crit}$ for evection resonance should be comparable to that of second-order ivection resonance.}

{Another possible explanation is that the deficiency of the planets near $P_B/P_{\rm prec}\sim 1$ is due to a suppression of planet formation, rather than removal of planets captured into ivection or evection resonance. i.e. it is possible that formation of planet (at a certain semi-major axis) is suppressed when the local precession period of planetesimals (due to, for instance, perturbation from the surrounding disk) is commensurate with $P_B$. This explanation is physically reasonable, since ivection resonance can drive planetesimals away from the midplane, thereby reducing their collision rate and suppressing planet formation.}

{Overall, more data (and further physical and statistical analysis) are required to verify whether the gap in $P_B/P_{\rm prec}$ distribution is physical, and, if the gap is indeed physical, how (and whether) ivection / evection resonance creates this gap.}

\section{Discussion}\label{sec:discussion}
\subsection{Comparison with evection resonance}
Ivection resonance is very similar to evection resonance in that they both originate from the commensurability between the precession of the planets and the orbital motion of the binary. However, there are a few important differences between them.

Table \ref{scalings} summarizes the key properties and scaling relations for ivection and evection resonances; properties for MMR are also given as reference.
As shown in Table \ref{scalings}, evection resonance is a second-order resonance (i.e. with resonant term $\propto e_i^2$), with scalings identical to second-order ivection resonance.
To quadrupole order of the binary perturbation potential, this is the only type of evection resonance; there is no evection counterpart of first-order ivection resonance or eccentric ivection resonance.

The direction of migration required for resonance capture is also different for ivection and evection resonance. For ivection resonance, resonance capture requires $|\dot\Omega|/n_B$ to increase, which requires planets to migrate convergently or inward. Meanwhile, evection resonance capture requires $|\dot\varpi_i|/n_B<0$ (note that the apsidal precession rate $\dot\varpi_i >0$), which requires planets to migrate divergently or outwards (see the examples in \citealt{ToumaSridhar15}).

\subsection{The effect of finite disk mass}
In our analysis, we have only considered the case when the precession of the planet is driven by another planet.
However, the gravitational perturbation from the disk can also drive the planet to precess.
Consider a single planet migrating in a disk; the gravitational perturbation of the disk mass drives the planet to precess with
\eq{
\frac{2\pi}{|\dot\Omega|} \sim \mu_D^{-1}P,\rm{~~with~~}\mu_D\equiv \frac{\Sigma_Da^2}{M_\star}.
}
Here $\Sigma_D$ is the disk surface density evaluated near the planet.
(The secular planet-disk interaction can be calculated more precisely with the method given in \citealt{Heppenheimer80}. For our purpose, a coarse scaling estimate is enough since we do not know the detailed disk profile in the first place.)
In other words, the disk acts like another planet with mass $\Sigma_Da^2$,
and all scaling relations in Table \ref{scalings} should still hold (with $\mu$ replaced by $\mu_D$).
Migration of the planet and evolution of the disk can change the precession rate, providing another mechanism of encountering (and capturing into) ivection resonances.

As we discussed in Section \ref{subsec:Tcrit}, one major factor that limits the possibility of ivection (and evection) resonance capture is that unless $(1-e_B)$ is small, $\mu$ has to be large ($\gtrsim$ a few $10^{-2}$) for $T_{\rm crit}$ to be sufficiently small. Since $\mu_D$ can be comparable to the disk to star mass ratio (which is often $\gtrsim 10^{-2}$), having $T_{\rm crit}$ below the actual migration timescale of the planet is much easier. 

One caveat is that the dispersal of disk tends to decrease the precession rate, while ivection resonance capture requires increasing precession rate.
Capture into ivection resonance becomes possible only when the migration of the planet tends to increase the precession rate, and the effect of migration overshadows the effect of disk dispersal.
Meanwhile, the dispersal of the disk makes the system more likely to cross evection resonance in the right direction for capture.

\subsection{A note on Lidov-Kozai mechanism}
Lidov-Kozai mechanism can also excite eccentricity and inclination of the planet. 
{However, when ivection or evection resonance is important, the nodal precession timescale of the planets ($\sim P_B$) is generally short enough to completely suppress mutual inclination perturbation due to Lidov-Kozai mechanism, which occurs at a timescale of $\sim (P_B^2/P) (1-e_B^2)^{3/2}$. The ratio between the two timescales is $\sim (P_B/P) (1-e_B^2)^{3/2}\sim \mu^{-1} (1-e_B^2)^{3/2}$, and Lidov-Kozai mechanism can be important only when the binary eccentricity is extremely high ($e_B>0.99$ for Jupiter-mass planets).}

{On the other hand, Lidov-Kozai mechanism may affect the eccentricity evolution of the planets if at least one of them have a relatively low apsidal precession rate, as discussed in \citet{Takeda2008}. This mechanism is irrelevant for systems like Kepler-108 where both planets are massive and undergo fast apsidal precession. However, it can become relevant for systems with more extreme mass ratios, where the apsidal precession of the more massive planet is slow.}


\section{Summary}\label{sec:summary}
We identify a new type of secular-orbital resonance, ivection resonance, that can excite the mutual inclination of planets in a multi-planet system (or a single-planet system with a relatively massive disk).
Ivection resonance happens when the nodal precession rate of the planet (due to perturbation from another planet or the protoplanetary disk) is commensurate with the orbital frequency of an external binary perturber.
We study several types of ivection resonances (Section \ref{sec:ivection}), including first and second-order ivection resonance (for circular binary perturber) and eccentric ivection resonance (for very eccentric binary). They share similar physical nature, but their strengths vary. Their properties are summarized in Table \ref{scalings}.

Capturing into an ivection resonance happens when the system encounters an ivection resonance with slowly increasing $|\dot\Omega|/n_B$. More precisely, the planet migration timescale needs to be shorter than a critical migration timescale $T_{\rm crit}$ (see Section \ref{subsec:Tcrit}), which is usually very large unless the planets are relatively massive or the binary is highly eccentric.

{After capturing into an ivection resonance, the mutual inclination increases as the system continues to migrate, while the eccentricities remain unaffected by ivection resonance.
If migration stops before the system encounters another resonance (such as a MMR), the system can stay resonant.
Otherwise, encountering with another resonance often disrupts ivection resonance. Depending on the strength of this second resonance, the system is either left at a mutually inclined non-resonant configuration or become dynamically unstable. In the latter case, the smaller planet may be ejected.}

{We propose ivection resonance as a promising mechanism for producing the orbital configuration of Kepler-108, a system hosting two planets with mild eccentricities but significant mutual inclination. We use simulations to reproduce the observed configuration of the system in Section \ref{sec:K108}.} The importance of ivection resonance in other exoplanet systems with relatively close binary companion is also investigated, with some indirect evidence (Section \ref{subsec:importance}) suggesting that ivection resonance may significantly affect the formation and / or evolution of such systems. However, given the large $T_{\rm crit}$ of most systems, our current theory of ivection resonance capture cannot account for such significant effect, and how ivection resonance affect the formation and evolution of planets perturbed by external binary companion should be investigated in future studies.

\acknowledgements

We thank the Institute for Advanced Study for hosting D.F. during which this work commenced, and Sean Mills for helpful discussion.
D.F. acknowledges support of grant NASA-NNX17AB93G through NASA's Exoplanet Research Program.

\bibliography{ms}

\newpage
\section*{Appendex}
\setcounter{section}{0}
\renewcommand{\thesection}{\Alph{section}}

\section{\red{Derivation of the effective Hamiltonian}}\label{a:Ham}
\subsection{Frame of reference and the evolution of mutual inclination}\label{a:frame}
We start by finding a frame of reference for orbital elements in which the evolution of the two planets' mutual inclination $I$ can be easily related to the evolution of their individual inclinations $I_i$ and longitudes of ascending node $\Omega_i$.

We can define a (generally time-dependent) ``standard" frame of reference, where the plane of reference is the invariable plane if one only takes into account the two planets (i.e. the plane normal to their total angular momentum), and the reference direction $\hat{\bs{x}}(t)$ in this plane is defined such that $d\hat{\bs{x}}/dt$ is perpendicular to the standard plane of reference.
A ``mutual" longitude of ascending node $\Omega$ can be defined as the outer planet's longitude of ascending node in this standard reference frame.
This standard reference frame rotates, mainly due to the precession of the invariable plane around the binary angular momentum.
For the systems we are interested in, the planet-planet coupling is much stronger than the planet-binary coupling, and the rotation of this frame is much slower than the nodal precession rate of the planets.

We then define a fixed frame of reference that coincide with our standard frame of reference at a certain time $t=t_0$. If we define $I_i$ and $\Omega_i$ with respect to this fixed frame of reference, $I$ and $\Omega$ at $t_0$ are simply given by $I = I_1+I_2$ and $\Omega=\Omega_1+\pi=\Omega_2$. Moreover, since $d\hat{\bs{x}}/dt$ is perpendicular to the plane of reference, it is easy to prove that at $t=t_0$,
\eq{
\frac{d}{dt}\left(\mathcal F(I)\cos\Omega\right) = \frac{d}{dt}\left(\sin I_2\cos\Omega_2-\sin I_1\cos\Omega_1\right),~~~\frac{d}{dt}\left(\mathcal F(I)\sin\Omega\right) = \frac{d}{dt}\left(\sin I_2\sin\Omega_2-\sin I_1\sin\Omega_1\right).\label{eq:Ii_to_I_1}
}
Here $\mathcal F(I)$ is a function of $I$, defined as $\mathcal F(I) = \sin I'_1 + \sin I'_2$ where $I'_i$ is the inclination of planet $i$ evaluated in the standard reference frame.\footnote{Mathematically, $I_i'$ can be calculated as a function of $I$ by solving $I'_1+I'_2=I$ and $\sin I'_1 = \beta \sin I'_2$, where $\beta$ is the (constant) ratio between the angular momenta of the two planets.}
Note that the $I'_i=I_i$ at $t=t_0$ and $\mathcal F(I) = I+\mathcal O(I^3)$ for small $I$.

Now consider the resonant angles, which are of form $\theta_i = -\Omega_i+\varphi(t)$ and $\theta = -\Omega+\varphi'(t)$. Here $\varphi(t)$ is some linear combination of the binary's orbital elements $(\Omega_B,\varpi_B,\lambda_B)$, evaluated with respect to the fixed frame defined above.
$\varphi'(t)$ is the same as $\varphi(t)$, except that it is evaluated with respect to the standard reference frame defined above.
Since the rotation of the standard reference frame is slow and non-resonant (i.e. at small $I$ the rate of rotation is insensitive to $\theta$), in this paper we ignore this rotation when evaluating the time derivatives of $\varphi(t)$ and $\varphi'(t)$ and assume
\eq{
\frac{d}{dt}\varphi(t) \approx \frac{d}{dt}\varphi'(t) \approx \frac{\partial \theta}{\partial \lambda_B} n_B.
}
In the remaining of this appendix we do not distinguish between $\varphi(t)$ and $\varphi'(t)$.
Under this approximation, \eqref{eq:Ii_to_I_1} suggests that at $t=t_0$,
\eq{
\frac{d}{dt}(\mathcal F(I)\cos\theta) = \frac{d}{dt}(\sin I_2\cos\theta_2-\sin I_1\cos\theta_1),~~~\frac{d}{dt}(\mathcal F(I)\sin\theta) = \frac{d}{dt}(\sin I_2\sin\theta_2-\sin I_1\sin\theta_1).\label{eq:Ii_to_I}
}
Note that here $I,\theta$ are defined with respect to the time-dependent standard reference frame and $I_i,\theta_i$ are defined with respect to the fixed reference frame. The above result holds only at $t=t_0$. (At a different time, we need to evaluate the RHS with respect to a different frame.)

In the calculations below, we will first find the equations of motion for $I_i$ and $\theta_i$, then use them to get the equations of motion for variables related to $I$ and $\theta$.

\subsection{Equations of motion for inclinations of individual planets}
The secular (averaged over planet orbits) Hamiltonian of the system can be written as
\eq{
H_{\rm sec} = \langle \Phi_{12} \rangle + \langle \Phi_{1B} \rangle_{\rm res} + \langle \Phi_{2B} \rangle_{\rm res}.
}
$\langle \Phi_{12} \rangle $ is the secular interaction between the two planets, and to fourth order in inclination \citep{MurrayDermott99}
\eq{
\langle \Phi_{12} \rangle = &-\frac{Gm_1m_2}{a_2}\left[(s_1^2+s_2^2)f_3 + (s_1^4+s_2^4)f_8 + s_1^2s_2^2f_9 + s_1s_2f_{14}\cos(\Omega_2-\Omega_1)\right.\\ &\left.+ s_1s_2(s_1^2+s_2^2)f_{16}\cos(\Omega_2-\Omega_1) + s_1^2s_2^2f_{26}\cos(2\Omega_2-2\Omega_1)\right].
}
Here $s_i = \sin(I_i/2)$ and $f_i$ are functions of $\alpha=a_1/a_2$ given in Appendix B of \citet{MurrayDermott99} evaluated at $j=0$. For simplicity, we assume that the planet orbits remain circular.

$\langle \Phi_{iB} \rangle$ corresponds to the perturbation from the binary and $\langle \Phi_{iB} \rangle_{\rm res}$ includes only the resonant terms in $\langle\Phi_{iB}\rangle$.
We neglect all non-resonant terms in $\langle \Phi_{iB} \rangle$, since the planets are tightly coupled and $|\langle \Phi_{12} \rangle| \gg |\langle \Phi_{iB} \rangle|$.
$\langle \Phi_{iB} \rangle_{\rm res}$ for different types of ivection resonances are given in \S\ref{sec:Phi_res} and Appendix \ref{a:eccentric}.

Consider the canonically conjugate variables $\mathcal G_i \equiv m_i\sqrt{GM_\star a_i}(2s_i^2)$ and $g_i\equiv -\Omega_i$. We want to make a canonical transform so that one of the conjugate variables is the resonant angle
\eq{
\theta_i \equiv \left\{\begin{array}{ll}
-2\lambda_B + 3\Omega_B-\Omega_i & \text{(1st order ivection)}\\
-\lambda_B + 2\Omega_B-\Omega_i & \text{(2nd order ivection)}\\
-j(\lambda_B-\varpi_B) - (\Omega_i-\Omega_B) & \text{(eccentric ivection)}\\
\end{array}\right.
}
In general, we can write $\theta_i = -\Omega_i + \varphi(t)$ [and $\theta=-\Omega+\varphi(t)$] with $\varphi(t)$ being a linear combination $\Omega_B,\varpi_B$ and $\lambda_B$, and $d\varphi(t)/dt\approx n_B~\partial \theta/\partial \lambda_B$.
Using a canonical transform with type-3 generating function $G_3(\theta_i,\mathcal G_i) = -\sum_{i=1,2}[\theta_i-\varphi(t)]\mathcal G_i$, we get a new set of canonical variables
$(\Theta_i, \theta_i)$ with $\Theta_i=\mathcal G_i$,
and the corresponding Hamiltonian is
\eq{
\tilde H_{\rm sec}(\Theta_i,\theta_i) = H_{\rm sec} + \frac{\partial G_3}{\partial t} = H_{\rm sec} + \frac{\partial \theta}{\partial \lambda_B}n_B\sum_{i=1,2}\Theta_i.
}
$(\Theta_i,\theta_i)$ can then be transformed to another pair of conjugate variables, $(\sqrt{2\Theta_i}\cos\theta_i,~\sqrt{2\Theta_i}\sin\theta_i)$. For convenience, we define a pair of dimensionless variables $(X_i,Y_i)$ as
\eq{
(X_i,Y_i) \equiv \frac{1}{m_i\sqrt{GM_\star a_i}}(\sqrt{2\Theta_i}\cos\theta_i,~\sqrt{2\Theta_i}\sin\theta_i) = (2s_i\cos\theta_i,2s_i\sin\theta_i).
}
The equations of motion for $(X_i,Y_i)$ are then
\eq{
\frac{dX_i}{dt} = -\frac{1}{m_i\sqrt{GM_\star a_i}}\frac{\partial\tilde H_{\rm sec}}{\partial Y_i},~~~
\frac{dY_i}{dt} = \frac{1}{m_i\sqrt{GM_\star a_i}}\frac{\partial\tilde H_{\rm sec}}{\partial X_i}.\label{eq:dXi_dt}
}
This gives the evolution of the inclinations of individual planets. Note that $(X_i,Y_i)$ can be defined with respect to any fixed reference frame.

\subsection{Equations of motion for mutual inclination}

Now let's consider the equations of motion for
\eq{
X\equiv \mathcal F(I)\cos\theta,~~~Y\equiv \mathcal F(I)\sin\theta.
}
$\mathcal F(I)$ is defined in \ref{a:frame} and $\mathcal F(I) = I+\mathcal O(I^3)$ for small $I$.
Using results from \ref{a:frame},
\eq{
\frac{d}{dt}X &= \frac{d}{dt}(\sin I_2\cos\theta_2-\sin I_1\cos\theta_1),\\
\frac{d}{dt}Y &= \frac{d}{dt}(\sin I_2\sin\theta_2-\sin I_1\sin\theta_1).
}
Since $\sin I_i = 2s_i (1-\frac 18 I_i^2) + \mathcal O(I_i^5)$, the RHS can be written in terms of $X_i, Y_i$ as (up to third order in inclination)
\eq{
\frac{d}{dt}X &= \sum_{i=1,2}(-1)^i \left\{\left[1-\frac{1}{8}(X_i^2+Y_i^2)\right]\frac{dX_i}{dt} - \frac{1}{4} \left(X_i^2\frac{dX_i}{dt}+X_iY_i\frac{dY_i}{dt}\right)\right\},\\
\frac{d}{dt}Y &= \sum_{i=1,2}(-1)^i \left\{\left[1-\frac{1}{8}(X_i^2+Y_i^2)\right]\frac{dY_i}{dt} - \frac{1}{4} \left(Y_i^2\frac{dY_i}{dt} + X_iY_i\frac{dX_i}{dt}\right)\right\}.
}
Here $d(X_i,Y_i)/dt$ are given by \eqref{eq:dXi_dt}.
Then we can rewrite the RHS in terms of $X,Y$ using the relation
\eq{
(X_1,Y_1) &= -\frac{\beta}{1+\beta}\left[1+\frac 18 \frac{\beta^2}{(1+\beta)^2}(X^2+Y^2)+\mathcal O(I^4)\right](X,Y),\\
(X_2,Y_2) &= \frac{1}{1+\beta}\left[1+\frac 18 \frac{1}{(1+\beta)^2}(X^2+Y^2) + \mathcal O(I^4) \right](X,Y).
}
Here $\beta = (m_2/m_1)(a_2/a_1)^{1/2}$ is the ratio between the angular momenta of the planets.

Below we summarize the explicit expressions for $d(X,Y)/dt$ in terms of $(X,Y)$.
We can split $d(X,Y)/dt$ into non-resonant terms (those associated with $\langle\Phi_{12}\rangle + \partial G_3/\partial t$) and resonant terms (those associated with $\langle\Phi_{iB}\rangle_{\rm res}$); we keep non-resonant terms up to third order in inclination, and only keep the lowest order resonant term because $|\Phi_{iB}|\ll |\Phi_{12}|$.

The non-resonant terms give (to third order in inclination)
\eq{
\left(\frac{dX}{dt}\right)_{\rm pp} &= \left\{-\frac{\partial\theta}{\partial\lambda_B}n_B + \dot\Omega_0\left[1+\frac 12 \left(\frac{f_8}{f_3}-\frac{\beta}{(1-\beta)^2}\right)(X^2+Y^2)\right]\right\}Y,\\
\left(\frac{dY}{dt}\right)_{\rm pp} &= \left\{\frac{\partial\theta}{\partial\lambda_B}n_B - \dot\Omega_0\left[1+\frac 12 \left(\frac{f_8}{f_3}-\frac{\beta}{(1-\beta)^2}\right)(X^2+Y^2)\right]\right\}X.
\label{eq:dX_sec}
}
The subscript ``pp" denotes $d(X,Y)/dt$ due to the non-resonant planet-planet interaction, and
\eq{
\dot\Omega_0\equiv \frac 12 f_3 n_2\frac{m_1(1+\beta)}{M_\star}.
}
To obtain the above results, we have used the following relations between $f_i$: $f_{14}=-2f_3, ~~f_{16}=f_3-4f_8$ and $f_9+f_{26}=6f_8-2f_3$.
These relations can be easily proved using the expressions for $f_i$ given in \citet{MurrayDermott99}, or using the physical argument that rotating the plane of reference should leave $\langle\Phi_{12}\rangle$ invariant.
The mutual precession rate due to planet-planet interaction can also be obtained from the above result: To second order in inclination, the mutual precession rate is
\eq{
\left(\frac{d\Omega}{dt}\right)_{\rm pp} = \dot\Omega_0\left[1+\frac 12 \left(\frac{f_8}{f_3}-\frac{\beta}{(1-\beta)^2}\right)I^2\right].
}

For the resonant terms, using $\langle\Phi_{iB}\rangle_{\rm res}$ from \S \ref{sec:Phi_res} and Appendix \ref{a:eccentric}, we find that (to lowest order in inclination),
\eal{
&\left(\frac{dX}{dt}\right)_{\rm res} = 0,~~~\left(\frac{dY}{dt}\right)_{\rm res} = -\frac{3}{4}\sin I_B\frac{1-\cos I_B}{2}\frac{q}{1+q}\frac{n_B^2(n_1-n_2)}{n_1n_2}~~~\text{for first order ivection resonance;}\label{eq:dX_res_1}\\
&\left(\frac{dY}{dt}\right)_{\rm res} = 0,~~~\left(\frac{dX}{dt}\right)_{\rm res} = -\frac 32 \left(\frac{1-\cos I_B}{2}\right)^2\frac{q}{1+q}\frac{n_B^2}{n_2}\frac{1+\beta n_2/n_1}{1+\beta}Y~~~\text{for second order ivection resonance;}\label{eq:dX_res_2}\\
&\left(\frac{dX}{dt}\right)_{\rm res} = 0,~~~\left(\frac{dY}{dt}\right)_{\rm res} = -\frac{3}{16\sqrt{2}}\frac{\sin(2I_B)}{(1-e_B)^{3/2}}\frac{q}{1+q}\frac{n_B^2(n_1-n_2)}{n_1n_2}~~~\text{for eccentric ivection resonance.}\label{eq:dX_res_3}
}
Here $q = M_B/M_\star$ is the binary mass ratio. Adding \eqref{eq:dX_res_1}, \eqref{eq:dX_res_2}, or \eqref{eq:dX_res_3} to \eqref{eq:dX_sec} then gives the equations of motion for $(X,Y)$.

\subsection{Constructing the effective Hamiltonian}
At small inclination, the equations of motion for $(X,Y)$ can also be described by an effective Hamiltonian given by
\eq{
H_{\rm eff} = \frac 12 \left(\frac{\partial\theta}{\partial\lambda_B}n_B-\dot\Omega_0\right)(X^2+Y^2) - \frac 18 \dot\Omega_0 \left(\frac{f_8}{f_3}-\frac{\beta}{(1-\beta)^2}\right)(X^2+Y^2)^2 + \Phi_{\rm eff,res},
}
where
\eq{
\Phi_{\rm eff,res} = \left\{\begin{array}{ll}
-\frac{3}{4}\sin I_B\frac{1-\cos I_B}{2}\frac{q}{1+q}\frac{n_B^2(n_1-n_2)}{n_1n_2}X & \text{for first order ivection resonance,}\\
\frac 34 \left(\frac{1-\cos I_B}{2}\right)^2\frac{q}{1+q}\frac{n_B^2}{n_2}\frac{1+\beta n_2/n_1}{1+\beta}Y^2 & \text{for second order ivection resonance,}\\
-\frac{3}{16\sqrt{2}}\frac{\sin(2I_B)}{(1-e_B)^{3/2}}\frac{q}{1+q}\frac{n_B^2(n_1-n_2)}{n_1n_2}X & \text{for eccentric ivection resonance.}
\end{array}\right.
}
It is easy to see that $H_{\rm eff}$ reproduces the correct equations of motion for $(X,Y)$, up to $\mathcal O(I^3)$ for the non-resonant (planet-planet) terms and to lowest nontrivial order for the resonant (planet-binary) term.
This effective Hamiltonian makes it easier to study the evolution of mutual inclination as the system passes the resonance.

We can scale this effective Hamiltonian into a dimensionless form, in order to compare it with other resonances (e.g. mean-motion resonance and evection resonance) with similar Hamiltonians.
We make the following transforms:
\eq{
\mathcal H = I_0^{-2}T_0H_{\rm eff},~~~(x,y) = (X,Y)/I_0,~~~\tau = t/T_0.
}
After the transform, the equations of motion become
\eq{
\frac{dx}{d\tau} = - \frac{\partial \mathcal H}{\partial y}, ~~~
\frac{dy}{d\tau} = \frac{\partial \mathcal H}{\partial x}.
}
We choose the scaling coefficients $I_0$ and $T_0$ such that $\mathcal H$ is of form $\eta(x^2+y^2)-(x^2+y^2)^2\pm x$ for first order and eccentric ivection resonance and of form $\eta(x^2+y^2)-(x^2+y^2)+y^2$ for second order ivection resonance. The signs of the resonant terms are chosen such that $I_0$ and $T_0$ are positive. $I_0$ and $T_0$ for different types of resonances have been summarized in \S\ref{sec:Ham}.

\section{\red{Resonant perturbation potential for eccentric ivection resonance}}\label{a:eccentric}
In this Appendix, we calculate the resonant perturbation potential $\langle \Phi_{iB}\rangle_{\rm res}$ for eccentric ivection resonance.

The secular perturbation potential due to the binary is
\eq{
\langle \Phi_{iB}\rangle = -GM_Bm_i\left\langle\frac{1}{|\bs r_B-\bs r_i|}-\frac{\bs r_B\cdot \bs r_i}{r_B^3}\right\rangle.\label{eq:Phi_iB_def}
}
In the calculation below, we assume that the planet orbits remain circular and $m_i/M_\star \to 0$. The second term in \eqref{eq:Phi_iB_def} therefore vanishes after averaging over the planet orbit. We then expand the potential in $a_i/r_B$ up to quadrupole order and expand in $\sin I_i$ up to first order. This gives
\eq{
\langle \Phi_{iB}\rangle = \frac 32 \frac{GM_Bm_ia_i^2}{r_B^5}\sin I_i\left(\sin\Omega_i x_Bz_B - \cos\Omega_i y_Bz_B\right) + \text{(terms independent of $I_i$ and $\Omega_i$)} + \text{(high-order terms)}.
}
Here $(x_B,y_B,z_B)$ is the location of the binary, defined in a coordinate system where $(\hat{\bs{x}},\hat{\bs{y}})$ spans the plane of reference and $\hat{\bs{x}}$ is aligned with the reference direction.


Now we Fourier expand this potential in terms of the mean anomaly $M=\lambda_B-\varpi_B$ (not to be confused with the stellar masses $M_\star$ and $M_B$):
\eq{
\frac 32 \frac{GM_Bm_ia_i^2}{r_B^5}\sin I_i\left(\sin\Omega_i x_Bz_B - \cos\Omega_i y_Bz_B\right) = \sum_{k=-\infty}^{\infty}A_k e^{ ikM}.
}
Here the Fourier amplitude $A_k$ is independent of $M$ (or $\lambda_B$).

In order to find $A_k$, we first compute the Fourier amplitudes of $(x'_B)^2/r_B^5$, $(y'_B)^2/r_B^5$ and $(x'_By'_B)/r_B^5$, where $x'_B\equiv r_B\cos M$ and $y'_B\equiv r_B\sin M$. For example, the Fourier amplitude of $(x'_B)^2/r_B^5$ will be
\eq{
\left(\frac{(x'_B)^2}{r_B^5}\right)_k = \frac{1}{2\pi}\int_{-\pi}^{\pi}dM e^{- ikM} \frac{(x'_B)^2}{r_B^5}.\label{eq:Fourier}
}
$x'_B, y'_B, r_B$ and $M$ can be related to the eccentric anomaly $E$ by
\eq{
r_B = a_B(1-e_B\cos E),~~x'_B = a_B(\cos E-e_B),~~~y'_B=a_B\sin E\sqrt{1-e_B^2},~~~M = E-e_B\sin E.
}
For small $(1-e_B)$, the integral for Fourier amplitude \eqref{eq:Fourier} is dominated by contribution from small $E$. To simplify the integral, we define a scaled variable $\zeta \equiv E/\sqrt{1-e_B}$, and push the limits of the integral from $\zeta = \pm \pi/\sqrt{1-e_B}$ to $\zeta = \pm\infty$. We also expand $x'_B,y'_B,r_B,\exp(-ikM)$ and $dM$ in terms of $\zeta$ and keep only the first two terms.\footnote{Here keeping only one term is not enough, because the two lowest order terms can often have similar coefficients. For example, $r_B \approx a_B(1-e_B)(1-\zeta^2/2)$. The coefficients for higher order terms, on the other hand, are smaller at least by a factor of $(1-e_B)$.}
This allows the Fourier amplitudes to be written as some coefficient (involving $a_B$ and $1-e_B$) times an integral that depends only on $\zeta$. Physically, this approximation corresponds to expanding the Fourier amplitude in terms of $(1-e_B)$ and keeping only the lowest order term.
After evaluating the integrals, we get
\eq{
\left(\frac{(x'_B)^2}{r_B^5}\right)_k \approx \left(\frac{(y'_B)^2}{r_B^5}\right)_k \approx \frac{1}{4\sqrt{2}}a_B^{-3}(1-e_B)^{-3/2},~~~\left(\frac{x'_B y'_B}{r_B^5}\right)_k= \mathcal O(a_B^{-3}(1-e_B)^{-1}) \approx 0.
}
Note that the above result is valid only for relatively small $k$: our expansion of the integrand is valid only when $|kM|\ll 1$ at $\zeta \sim 1$, and this requires $k\ll (1-e_B)^{-3/2}$.
We can then obtain $A_k$ by relating $(x_B,y_B,z_B)$ to $(x'_B,y'_B)$ via a rotation. To lowest order in $(1-e_B)$,
\eq{
A_k = -\frac{3}{16\sqrt{2}}\frac{GM_Bm_ia_i^2}{a_B^3(1-e_B)^{3/2}}\sin I_i \sin(2I_B)\cos(\Omega_i-\Omega_B).
}
Therefore, for an eccentric ivection resonance with $\dot\Omega\approx -jn_B$ ($j$ is a positive integer, since $\dot\Omega<0$), the resonant term in $\langle\Phi_{iB}\rangle$ is given by
\eq{
\langle \Phi_{iB}\rangle_{\rm res} = -\frac{3}{16\sqrt{2}}\frac{GM_Bm_ia_i^2}{a_B^3(1-e_B)^{3/2}}\sin I_i \sin(2I_B)\cos\theta_i.
}
where $\theta_i = -(\Omega_i-\Omega_B)-j(\lambda_B-\varpi_B)$ is the resonant angle.

\section{Testing the gap in $P_B/P_{\rm prec}$ distribution}\label{a:test}
In Figure \ref{Prat_distribution} the distribution of estimated $P_B/P_{\rm prec}$ for systems with single observed planet appears to have a gap near $P_B/P_{\rm prec}=1$, while the same distribution for system with multiple observed planets does not.
Here we discuss the statistical significance of this gap, and whether it is an artifact due to binning.

Consider the probability of observing a system with certain $P_B/P_{\rm prec}$. If this probability distribution does not contain any gap associated with ivection / evection resonance, it should be approximately be given by ``smearing out" the observed distribution. i.e. we can approximate the actual distribution by replacing each observed system with a normal distribution [in $\log_{10}(P_B/P_{\rm crit})$] centered at the observed value with a certain standard deviation $\sigma$. Here we choose $\sigma=1$ dex, which is about the smallest $\sigma$ for which the smeared distribution is no longer bimodal. This smeared distribution, after a normalization, gives our estimated probability distribution of observing a system with certain $P_B/P_{\rm prec}$ if there is no gap.

We then consider whether the observed data is consistent with this estimated gap-less distribution.
For each bin in the histogram, we calculate $p_\leq$, the probability that a sample drawn from the estimated distribution with the same sample size as observation contains $\leq N_{\rm obs}$ system in this bin, where $N_{\rm obs}$ is the number of observed systems in this bin.

\begin{figure}
\centering
\includegraphics[width=.5\textwidth]{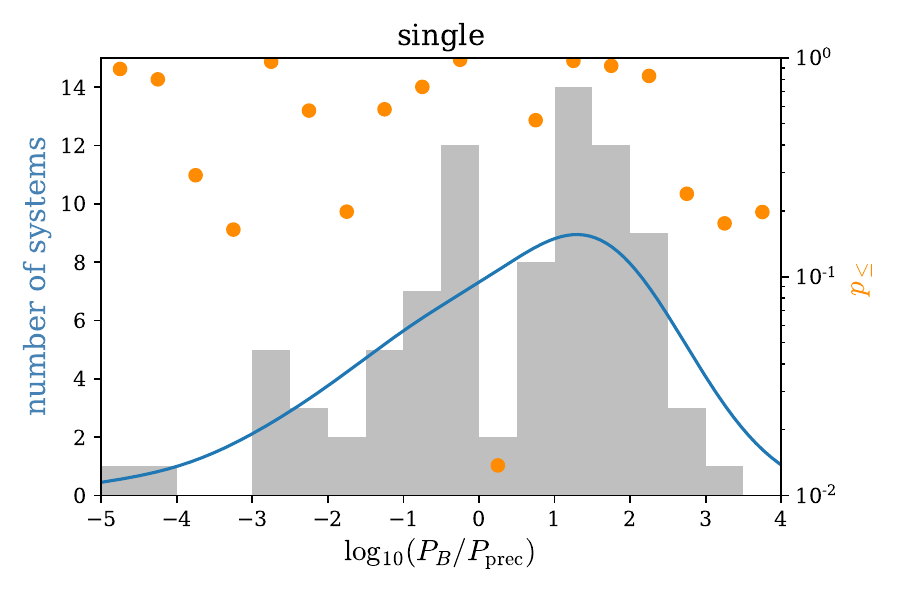}\includegraphics[width=.5\textwidth]{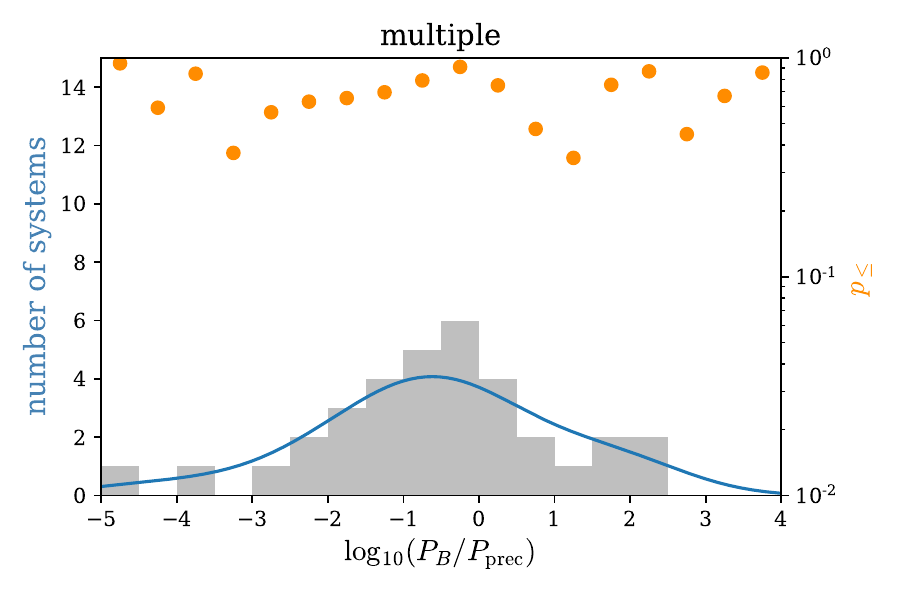}
\caption{The distribution of (estimated) $P_B/P_{\rm prec}$ for observed system (grey histogram), estimated distribution of $P_B/P_{\rm prec}$ assuming there is no gap (blue line; obtained by smearing the observation, see text for detail), and $p_\leq$ for each bin (orange dots; see definition of $p_\leq$ in text). Smaller $p_\leq$ means more significant inconsistency with the gap-less distribution.}
\label{Prat_distribution_test}
\end{figure}

We perform this analysis for $P_B/P_{\rm prec}$ distribution of systems with single or multiple observed planets, and the results are shown in Figure \ref{Prat_distribution_test}. For systems with single observed planet, $p_\leq \approx 0.015$ for the bin between 0 and 0.5 dex.
{We choose our null hypothesis to be that none of the 4 bins covering $P_B/P_{\rm prec}$ between -1 and 1 dex shows a gap.\footnote{The uncertainty of the gap location is mainly because our estimate may contain systematic error of order unity.} and apply a Bonferroni correction.
This gives a significance level of 0.06, which is not (although close to) statistically significant.
For systems with multiple observed planets, there is neither visible nor statistically significant gap, probably due to the smaller sample size.}

We also perform the same analysis for $P_B$ and $P_{\rm prec}$. We find that $p_\leq$ is always large in their distributions, suggesting that the gap in the left panel of Figure \ref{Prat_distribution_test}, if not a coincidence, should be due to effect(s) directly associated with $P_B/P_{\rm prec}$ such as ivection / evection resonance.

Additionally, we address the concern whether the gap in the left panel of Figure \ref{Prat_distribution_test} is due to a particular binning choice, by varying the number of bins.
This changes both bin size and bin location at the gap. We find that for a wide range of bin numbers (15 - 25 bins from -5 dex to 4 dex, while Figure \ref{Prat_distribution_test} has 18), the minimum $p_\leq$ is $\lesssim 0.015$ more than half of the time (since the gap width may be comparable to the bin size, it is natural for the gap to appear less significant when it is between two bins), confirming that this gap is not just due to the binning.

\end{document}